\documentclass[pra,reprint]{revtex4-1}
\usepackage{graphicx}
\usepackage{appendix}
\usepackage{braket}
\usepackage{rotating}
\usepackage{simplewick}
\usepackage{amssymb}
\usepackage{amsmath}
\usepackage{txfonts}
\usepackage{bm}
\usepackage{color}
\usepackage{multirow}
\usepackage{booktabs}
\usepackage{dcolumn}

\begin{document}

\title{Nonvanishing quadrature derivatives in the analytical gradients of density functional energies in crystals and helices}

\author{So Hirata}
\email{sohirata@illinois.edu}
\affiliation{Department of Chemistry, University of Illinois at Urbana-Champaign, Urbana, Illinois 61801, USA}

\date{\today}

\begin{abstract}
It is shown that the quadrature derivatives in some analytical gradients of energies evaluated 
with a multi-centre radial-angular grid do not vanish even in the limit of an infinitely dense grid, causing severe errors when neglected. 
The gradients in question are those with respect to a lattice constant of a crystal 
or to the helical angle of a chain with screw axis symmetry. This is in contrast with 
the quadrature derivatives in atomic gradients, which can be made arbitrarily small by grid extension. 
The disparate behaviour is traced to whether the grid points depend on the coordinate with respect to which the derivative of energy is taken.
Whereas the nonvanishing quadrature derivative in the lattice-constant gradient is identified 
as the surface integral arising from an expanding integration domain, 
the analytical origin of the nonvanishing quadrature derivative in the helical-angle gradient remains unknown.
\end{abstract}

\maketitle 

%\begin{figure}
%\includegraphics[width=0.5\columnwidth]{pe_170.jpg}
%\caption{TOC graphics}
%\end{figure}

\section{Introduction} 

The analytical energy gradient method of density functional theory (DFT) \cite{Parr1989,Ziegler1991}
is one of the most heavily used computational  machineries in chemistry. 
It serves as essential infrastructure of chemical research, upon which its accuracy, efficiency, and stability
have great impact.

The method is based on Pulay's formalism \cite{Pulay1969} of the analytical energy gradients for Hartree--Fock (HF) theory.
In DFT, an additional term emerges from the exchange-correlation (XC) energy, which is often evaluated by quadrature using  
Becke's interlocking multi-atom-centre grid \cite{Becke1988} (see also Satoko \cite{Satoko}).
Its formulation was first reported by Delley \cite{Delley1991}, who not only established a method for computing these XC energy gradients 
but also pointed out the existence of a correction due to the variations in grid coordinates and weights.
Delley and coworkers \cite{Delley1991,Baker1994} deemed this correction negligible in atomic gradients. 
Johnson, Gill, and Pople \cite{Johnson1993}\ reported an implementation fully accounting 
for these quadrature derivatives. Johnson and Frisch \cite{JohnsonCPL1993}, Malagoli and Baker \cite{Malagoli2003}, and more recently Shang and Yang \cite{Shang2020}
all concluded that the quadrature derivatives are nonnegligible in vibrational frequencies, although
they are generally small and can be made arbitrarily smaller by increasing grid size.
%Since the quadrature derivatives are inexpensive to evaluate, there is no reason not to include them.

The objective of this article is to point out surprising cases where the quadrature derivatives 
are large and nonvanishing even in the limit of an infinitely dense grid, causing severe errors when they are neglected.
This problem manifests itself in the atomic-basis-set DFT for crystals and helices \cite{Hirata1997,Hirata1998}.

\section{Quadrature derivatives} 

Let us take an infinite helical polymer (screw axis along $x$ with lattice constant or ``rise'' $a$ and helical angle or ``twist'' $\theta$ \cite{Hauser}) as our example \cite{Hirata1997,Hirata1998}.
The atomic positions in the $n$th helical repeat unit cell (hereafter simply ``cell'') are rotated anticlockwise by 
$n\theta$. Therefore, the coordinates of the $I$th atom in the $n$th cell are related to those in the zeroth cell by
\begin{eqnarray}
X_{I(n)} &=& X_{I(0)}+na, \label{X} \\
Y_{I(n)} &=& Y_{I(0)} \cos n\theta - Z_{I(0)} \sin n\theta, \label{Y} \\
Z_{I(n)} &=& Y_{I(0)} \sin n\theta + Z_{I(0)} \cos n\theta. \label{Z}
\end{eqnarray}
The formulation for this system encompasses molecules (where nonvanishing quadrature derivatives do not occur),
linear (nonhelical) polymers, and two- and three-dimensional crystals. Spin-restricted hybrid DFT formalism \cite{Teramae1983,Teramea1984,Hirata1997,Hirata1998} for the energy per cell is given in Appendix \ref{A}. For a more complete formalism of the analytical energy gradient with respect to $\theta$,
see Appendix \ref{B}. Here, we focus our attention on the quadrature derivatives in the XC energy gradients. Atomic units are used throughout this article unless otherwise stated.

The XC energy per cell is defined by
\begin{eqnarray}
E_{\text{XC}} &=& \int_0^a dx \iint dydz\, f_{\text{XC}}(\rho_\alpha,\rho_\beta,\gamma_{\alpha\alpha},\gamma_{\alpha\beta},\gamma_{\beta\beta}), \label{exc}
\end{eqnarray}
where $\rho$'s are the electron spin densities and $\gamma$'s are the spin density gradient invariants \cite{Johnson1993}.
This integral is usually evaluated by quadrature with Becke's interlocking multi-atom-centre grid \cite{Becke1988} (see also Satoko \cite{Satoko}) as
\begin{eqnarray}
E_{\text{XC}} &\approx& \sum_{I}\sum_{g(I)}  w_I(\bm{r}_{g(I)}) f_{\text{XC}}(\bm{r}_{g(I)}) , \label{exc_grid}
\end{eqnarray}
where $I$ runs over all atoms in the zeroth cell, $g(I)$ over all atom-centre grid points of the $I$th atom,
$\bm{r}_{g(I)}$ is the coordinates of the $g(I)$th grid point,
and $w_I(\bm{r}_{g(I)})$ is the quadrature weight.
It should be stressed that the integration domain is always the zeroth cell (although its boundaries are ``fuzzy'' \cite{Becke1988})
and the grid points and weights are defined only within that cell.

Since the XC energy depends on density $\rho$, atom-centre basis functions, and atom-centre grid,
its analytical gradient with respect to a general coordinate $\xi$ also consists of three terms:
\begin{eqnarray}
{ \frac{\partial E_{\text{XC}}}{\partial \xi}  }
= { \frac{\partial E_{\text{XC}}}{\partial \xi}  }^{(b)} + { \frac{\partial E_{\text{XC}}}{\partial \xi}  }^{(d)} 
+ { \frac{\partial E_{\text{XC}}}{\partial \xi} }^{(q)}. \label{bdq}
\end{eqnarray}
The first term with superscript `($b$)' involves the {\it direct} derivatives of the basis functions (not through variation in grid coordinates).
The second term with `($d$)' comes from the derivatives of the density matrix elements, which are
absorbed in the so-called Pulay force \cite{Pulay1969}. See Appendix \ref{B} for the first two terms.
The third term with `($q$)' is the quadrature derivatives, which consist of the derivatives of grid weights and
the derivatives with respect to grid coordinates. 

\subsection{In-phase atomic gradients} 

If $\xi$ is an in-phase atomic coordinate, the sum of the first two terms of Eq.\ (\ref{bdq}) (evaluated by the well-known method \cite{Delley1991,Johnson1993,Hirata1997};
see Appendix \ref{B}) is nearly equal to 
the total XC energy gradient \cite{Delley1991} for a typical grid size. Small errors due to the neglect of the quadrature derivatives (the third term) are 
on the order of $10^{-5}$ a.u.\ for a grid used in the preceding energy calculation, which are nevertheless noticeable as a small violation
of the translational invariance \cite{Johnson1993}. 
The quadrature derivatives erase these small errors and restore the translational invariance at near machine precision (see Sec.\ \ref{Numerical}).

Taking the $x$ coordinate of atom $I$ in the zeroth cell for $\xi$, the quadrature derivative is written as 
\begin{eqnarray}
{ \frac{\partial E_{\text{XC}}}{\partial X_{I(0)}}  }^{(q)}
&=& \sum_{J}^{\text{nuc.}}\sum_{g(J)}\frac{\partial w_J(\bm{r}_{g(J)}) }{\partial X_{I(0)}}  f_{\text{XC}}(\bm{r}_{g(J)}) \nonumber\\&&
+ \sum_{J}^{\text{nuc.}}\sum_{g(J)}w_N(\bm{r}_{g(J)}) { \frac{\partial  f_{\text{XC}}(\bm{r}_{g(J)})  }{\partial X_{I(0)}} }^{(q)} \label{xa1} \\
&=& \sum_{J}^{\text{nuc.}}\sum_{g(J)}\frac{\partial w_J(\bm{r}_{g(J)}) }{\partial X_{I(0)}}  f_{\text{XC}}(\bm{r}_{g(J)}) \nonumber\\&&
 + \sum_{J}^{\text{nuc.}}\sum_{g(J)}w_J(\bm{r}_{g(J)})  \frac{\partial \bm{r}_{g(J)} }{\partial X_{I(0)}} \frac{\partial  f_{\text{XC}}(\bm{r}_{g(J)})  }{\partial \bm{r}_{g(J)}} \\
%&=& \sum_{N}\sum_{g(N)}\frac{\partial w_N(\bm{r}_{g(N)}) }{\partial x_{A}^{(0)}}  f_{\text{XC}}( \bm{r}_{g(N)}) \nonumber\\&&
%+ \sum_{N}\sum_{g(N)}w_N(\bm{r}_{g(N)}) \delta_{NA} \frac{\partial  f_{\text{XC}}( \bm{r}_{g(N)})  }{\partial x_{g(N)}} \\
&=& \sum_{J}^{\text{nuc.}}\sum_{g(J)}\frac{\partial w_J(\bm{r}_{g(J)}) }{\partial X_{I(0)}}  f_{\text{XC}}(\bm{r}_{g(J)}) \nonumber\\&&
+ \sum_{g(I)}w_I(\bm{r}_{g(I)}) { \frac{\partial  f_{\text{XC}}(\bm{r}_{g(I)})  }{\partial x} }  .
\end{eqnarray}
The first and second terms in the right-hand sides are respectively the derivative of grid weights and the derivative 
with respect to grid coordinates \cite{Delley1991}. They have a large and nearly equal magnitude with opposite signs, cancelling each other nearly exactly.
This is why the quadrature derivatives are negligibly small in atomic gradients.

The programmable expression of the first term can be found in Ref.\ \cite{Johnson1993}. It is however more easily implemented by directly
applying the chain rule of differentiation to the code that calculates the quadrature weights. The second term is 
also easily programmed as a  minor modification to the code for $x$ atomic gradients (see Appendix \ref{B}).

\subsection{Lattice-constant gradient} 

Next, let us consider the gradient with respect to lattice constant $a$. 
Generally, such a gradient should be obtained as a by-product of atomic gradients along $x$ axis 
by virtue of the following relationship \cite{Teramae1983}:
\begin{eqnarray}
\frac{\partial}{\partial a} &=&\sum_{n=-S}^S \sum_{I}   \frac{\partial X_{I(n)}}{\partial a} \frac{\partial}{\partial X_{I(n)}} \\
&=&  \sum_{n=-S}^S \sum_{I}  n\, \frac{\partial}{\partial X_{I(n)}}, \label{da}
\end{eqnarray}
where $n$ runs over the same range of cell indexes adopted for the lattice sums \cite{Delhalle} in the preceding energy calculation, $I$ over all atoms in a cell, and Eq.\ (\ref{X}) 
was used. This gives the correct lattice-constant gradient of the HF energy, as pointed out by Teramae {\it et al.}\ \cite{Teramae1983}.

We found \cite{Hirata1997} that the above formula does not work for the atomic-basis-set DFT. This is because 
the integration domain of the XC energy depends on $a$ [see Eq.\ (\ref{exc})] and the formula ignores such dependence \cite{Hirata1997}.
Instead of Eq.\ (\ref{bdq}), we had to begin with the following breakdown of the lattice-constant gradient:
\begin{eqnarray}
{ \frac{\partial E_{\text{XC}}}{\partial a}  }
= { \frac{\partial E_{\text{XC}}}{\partial a}  }^{(b)} + { \frac{\partial E_{\text{XC}}}{\partial a}  }^{(d)} 
+ { \frac{\partial E_{\text{XC}}}{\partial a} }^{(s)}, \label{bds}
\end{eqnarray}
where the last term with superscript `$(s)$' stands for a surface integral arising from the explicit dependence of 
the integration domain on $a$. Specifically, it is the integral of the XC function over the cell boundary at $x=a$,
\begin{eqnarray}
{ \frac{\partial E_{\text{XC}}}{\partial a }  }^{(s)} &=& { \frac{\partial}{\partial a }  }^{(s)} \int_0^a dx \iint dydz\, f_{\text{XC}}  \\
&=&   \iint dydz\, f_{\text{XC}}(x=a) , \label{surface}
\end{eqnarray}
which is substantial and should not be neglected. In principle, this can be evaluated by two-dimensional quadrature, but 
we used the following mathematical trick to convert it into a volume integral and evaluated it with the same three-dimensional quadrature of Becke \cite{Hirata1997}:
\begin{eqnarray}
&& \iint dydz\, f_{\text{XC}}(x=a) \nonumber\\
&& =  \iint dydz\, \frac{x \, f_{\text{XC}}(x=0,a) }{a} \\
&& = \int_0^a dx \iint dydz \, \frac{\partial}{\partial x} \frac{x \, f_{\text{XC}}(x=0,a) }{a} \\
&& = \frac{E_{\text{XC}}}{a} + \sum_{I}\sum_{g(I)}w_I(\bm{r}_{g(I)}) \frac{x_{g(I)}}{a} { \frac{\partial  f_{\text{XC}}( \bm{r}_{g(I)})  }{\partial x} }^{(b)},
\end{eqnarray}
where the penultimate equality used Gauss' theorem. In this article, we call this term the surface-integral correction, although it is evaluated as the volume integral.

Comparing Eqs.\ (\ref{bdq}) and (\ref{bds}), we can identify the surface integral with the quadrature derivative (albeit not identically but closely; see Sec.\ \ref{Numerical}), which is nonvanishing 
regardless of grid size. 
The quadrature derivative with respect to $a$ is written as
\begin{eqnarray}
{ \frac{\partial E_{\text{XC}}}{\partial a }  }^{(q)}
&=& \sum_{I}\sum_{g(I)}\frac{\partial w_I(\bm{r}_{g(I)}) }{\partial a}  f_{\text{XC}}( \bm{r}_{g(I)}) \nonumber\\&&
+ \sum_{I}\sum_{g(I)}w_I(\bm{r}_{g(I)}) { \frac{\partial  f_{\text{XC}}( \bm{r}_{g(I)})  }{\partial a} }^{(q)} \\
&=& \sum_{I}\sum_{g(I)}\frac{\partial w_I(\bm{r}_{g(I)}) }{\partial a}  f_{\text{XC}}( \bm{r}_{g(I)}) \nonumber\\&&
+ \sum_{I}\sum_{g(I)}w_I(\bm{r}_{g(I)})  \frac{\partial \bm{r}_{g(I)} }{\partial a} \frac{\partial  f_{\text{XC}}( \bm{r}_{g(I)})  }{\partial \bm{r}_{g(I)}} \\
&=& \sum_{I}\sum_{g(I)}\frac{\partial w_I(\bm{r}_{g(I)}) }{\partial a}  f_{\text{XC}}( \bm{r}_{g(I)}).
\end{eqnarray}
Since no grid points in the zeroth cell move with $a$ (recall $E_\text{XC}$ is an integral within the volume of the zeroth cell), 
the second term (the derivatives with respect to grid coordinates) vanishes identically. The first term survives because Becke's fuzzy cell weights for grid points in the zeroth cell
are influenced by grid points of nearby atoms, which may move with $a$. 
Consequently, near exact cancellation of the two terms does not occur in the quadrature derivatives with respect to lattice constant.
The quadrature derivative in the lattice-constant gradient  
remains large and should not be neglected. 

\subsection{Helical-angle gradient} 

Invoking the same strategy embodied by Eq.\ (\ref{da}), we may be led to an identity,
\begin{eqnarray}
\frac{\partial}{\partial \theta} &\stackrel{?}{=}&\sum_{n=-S}^S  \sum_{I} 
 \left\{ \frac{\partial Y_{I(n)}}{\partial \theta}\frac{\partial Y_{I(0)}}{\partial Y_{I(n)}}\frac{\partial}{\partial Y_{I(0)}} 
 + \frac{\partial Y_{I(n)}}{\partial \theta}\frac{\partial Z_{I(0)}}{\partial Y_{I(n)}}\frac{\partial}{\partial Z_{I(0)}} \right. \nonumber\\
&& \left. + \frac{\partial Z_{I(n)}}{\partial \theta}\frac{\partial Y_{I(0)}}{\partial Z_{I(n)}}\frac{\partial}{\partial Y_{I(0)}}
+\frac{\partial Z_{I(n)}}{\partial \theta}\frac{\partial Z_{I(0)}}{\partial Z_{I(n)}}\frac{\partial}{\partial Z_{I(0)}}   \right\} \\
&\stackrel{?}{=}& \sum_{n=-S}^S \sum_{I}   n \left\{ Y_{I(0)} \frac{\partial}{\partial Z_{I(0)}} -Z_{I(0)} \frac{\partial}{\partial Y_{I(0)}}  \right\}. \label{wrong}
\end{eqnarray}
This appears to suggest that the helical-angle gradient should be obtainable as a by-product of atomic gradients along $y$ and $z$ axes.
This turns out to be false even at the HF level. What is erroneously neglected here [but not in Eq.\ (\ref{da})] is the fact that 
the basis functions not only shift their centres but also rotate with the helix (and only with these translating-rotating basis functions
does the Bloch theorem hold). See Appendix \ref{B} for the helical-angle gradient formalism 
that takes into account all $\theta$ dependencies including the basis function rotation.

The quadrature derivative with respect to $\theta$ reads
\begin{eqnarray}
{ \frac{\partial E_{\text{XC}}}{\partial \theta}  }^{(q)}
&=& \sum_{I}\sum_{g(I)}\frac{\partial w_I(\bm{r}_{g(I)}) }{\partial \theta}  f_{\text{XC}}( \bm{r}_{g(I)}) \nonumber\\&&
+ \sum_{I}\sum_{g(I)}w_I(\bm{r}_{g(I)}) { \frac{\partial  f_{\text{XC}}( \bm{r}_{g(I)})  }{\partial \theta} }^{(q)} \\
&=& \sum_{I}\sum_{g(I)}\frac{\partial w_I(\bm{r}_{g(I)}) }{\partial \theta}  f_{\text{XC}}( \bm{r}_{g(I)}) \nonumber\\&&
+ \sum_{I}\sum_{g(I)}w_I(\bm{r}_{g(I)})  \frac{\partial \bm{r}_{g(I)} }{\partial \theta} \frac{\partial  f_{\text{XC}}(\bm{r}_{g(I)})  }{\partial \bm{r}_{g(I)}} \\
&=& \sum_{I}\sum_{g(I)}\frac{\partial w_I(\bm{r}_{g(I)}) }{\partial \theta}  f_{\text{XC}}( \bm{r}_{g(I)}), \label{theta}
\end{eqnarray}
which also lacks the contribution arising
from the derivatives with respect to grid coordinates because the grid is defined only in the zeroth cell and its coordinates do not depend on $\theta$
(but its weights can vary with $\theta$). 
It is therefore nonnegligible no matter how dense the grid is. 

The nonvanishing quadrature derivative implies that the helical-angle gradient 
also has an additional term analogous to the surface integral in the lattice-constant gradient.
This is not self-evident because the integration domain (i.e., the cell volume) does not expand or contract with $\theta$, but it is only twisted relative to the adjacent domains.
We have been unable to identify its mathematical form or analytical origin. 
Viewed chemically, the variation in $\theta$ has the similar effect of lengthening or shortening the chemical bond across the boundary between the zeroth and first cells
as does the variation in $a$. Therefore, if we take as the integration domain a tube enclosing all chemical bonds that is terminated by a plane perpendicular to the bond across
the cell boundary, the helical-angle gradient should have a surface integral on the terminal plane, 
which is analogous to the surface integral in the lattice-constant gradient. 
Viewed mathematically, however, the concept of chemical bond seems incongruous and 
the origin of the additional term remains a mystery, although
its numerical value can be easily determined as the quadrature derivative. In this sense, 
taking into account the quadrature derivatives of atom-centre grids seems to be the most general and reliable 
method of formulating analytical derivatives even when a different numerical integration method is being employed.

\section{Numerical Analysis\label{Numerical}}

%\begin{squeezetable}
\begin{table*}
\caption{Energy gradients of helical polyethylene computed by B3LYP/6-31G (using VWN5). Short-range ($S$) and long-range ($L$) lattice sum cutoffs of $10$ and $30$ were 
used. The number of $k$ points ($K$) in the reciprocal unit cell was $100$. \label{6-31G}}
\begin{ruledtabular}
\begin{tabular}{lcrrrrrr}
Method\tablenotemark[1] & Grid\tablenotemark[2] & \multicolumn{1}{c}{$\partial E / \partial X_{\text{H}_1}$} & \multicolumn{1}{c}{$\partial E / \partial X_{\text{C}}$} & \multicolumn{1}{c}{$\partial E / \partial X_{\text{H}_2}$} & \multicolumn{1}{c}{$|\sum\partial E / \partial X|$\tablenotemark[3]} & \multicolumn{1}{c}{$\partial E / \partial a$} & \multicolumn{1}{c}{$\partial E / \partial \theta$}  \\ \hline
No quad. derivatives & 8656 & $+0.0017821$ & $-0.0007924$ & $-0.0009717$ & $2\times10^{-5}$ & $+0.3855667$ & $-0.0034748$ \\
+ Surface integral & 8656 & & & & & $-0.0335962$ & \\
Quad. derivatives& 8656 & $+0.0017118$ & $-0.0007672$ & $-0.0009447$ & $< 10^{-10}$ & $-0.0335790$ & $-0.0102510$ \\
Finite differences & 8656 & $+0.0017118$ & $-0.0007672$ & $-0.0009447$ & $4\times10^{-8}$ & $-0.0335790$ & $-0.0102507$  \\  \hline
No quad. derivatives & $25\times86$ & $+0.0018605$ & $-0.0008917$ & $-0.0010068$ & $4\times10^{-5}$ & $+0.3865811$ & $-0.0035099$  \\
+ Surface integral & $25\times86$ & & & & & $-0.0340325$ & \\
Quad. derivatives& $25\times86$ & $+0.0012946$ & $-0.0006241$ & $-0.0006705$ & $< 10^{-10}$ & $-0.0338825$ & $-0.0102618$ \\
Finite differences &  $25\times86$ & $+0.0012946$ & $-0.0006241$ & $-0.0006705$ & $3\times10^{-8}$ & $-0.0338826$ & $-0.0102616$  \\  \hline
No quad. derivatives & $50\times194$ & $+0.0017550$ & $-0.0007963$ & $-0.0009652$ & $6\times10^{-6}$ & $+0.3856114$ & $-0.0034914$  \\ 
+ Surface integral & $50\times194$ & & & & & $-0.0335432$ & \\
Quad. derivatives& $50\times194$ & $+0.0019589$ & $-0.0009519$ & $-0.0010070$ & $1 \times 10^{-10}$ & $-0.0334430$ & $-0.0102387$ \\  \hline
No quad. derivatives & $100\times302$ & $+0.0017793$ & $-0.0007983$ & $-0.0009699$ & $1\times10^{-5}$ & $+0.3855765$ & $-0.0034752$  \\
+ Surface integral & $100\times302$ & & & & & $-0.0335858$ & \\
Quad. derivatives& $100\times302$ & $+0.0017292$ & $-0.0007612$ & $-0.0009680$ & $< 10^{-10}$ & $-0.0335834$ & $-0.0102532$ \\  \hline
No quad. derivatives & $200\times770$ & $+0.0017767$ & $-0.0008059$ & $-0.0009709$ & $6\times10^{-8}$ & $+0.3855824$ & $-0.0034764$  \\
+ Surface integral & $200\times770$ & & & & & $-0.0335860$ & \\
Quad. derivatives& $200\times770$ & $+0.0017775$ & $-0.0008062$ & $-0.0009712$ & $< 10^{-10}$ & $-0.0335857$ & $-0.0102490$ \\  \hline
No quad. derivatives & $300\times974$ & $+0.0017768$ & $-0.0008059$ & $-0.0009709$ & $5\times10^{-9}$ & $+0.3855825$ & $-0.0034763$  \\
+ Surface integral & $300\times974$ & & & & & $-0.0335860$ & \\
Quad. derivatives& $300\times974$ & $+0.0017766$ & $-0.0008058$ & $-0.0009708$ & $< 10^{-10}$ & $-0.0335860$ & $-0.0102493$ \\
\end{tabular} 
\end{ruledtabular}
\tablenotetext[1]{``No quad.\ derivatives'' means the analytical gradient calculations without the quadrature derivatives or surface-integral correction.
``+ Surface integral'' is the ``No quad.\ derivatives'' calculation plus the surface-integral correction for the lattice-constant gradient. 
``Quad. derivatives'' stands for the analytical gradient calculations including the quadrature derivatives.
``Finite differences'' refers to the numerical gradient calculations using a symmetric two-point finite-difference formula with a displacement of $\pm0.001$ bohr or $\pm0.01^\circ$.}
\tablenotetext[2]{The number of Gauss--Chebyshev radial grid points per atom $\times$ the number of Lebedev angular grid points per atom. `8656' refers to a mixed grid of $25 \times 302+13 \times 50+12 \times 38$.}
\tablenotetext[3]{The absolute value of the sum of the three in-phase atomic gradients along $x$ axis, which reports the deviation from the perfect translational invariance.}
\end{table*}
%\end{squeezetable}

%\begin{squeezetable}
\begin{table*}
\caption{Energy gradients of helical polyethylene computed by B3LYP/cc-pVDZ (using VWN5 and spherical $d$ orbitals). Short-range ($S$) and long-range ($L$) lattice sum cutoffs of $10$ and $30$ were 
used. The number of $k$ points ($K$) in the reciprocal unit cell was $100$.  \label{cc-pVDZ}}
\begin{ruledtabular}
\begin{tabular}{lcrrrrrr}
Method\tablenotemark[1] & Grid\tablenotemark[2] & \multicolumn{1}{c}{$\partial E / \partial X_{\text{H}_1}$} & \multicolumn{1}{c}{$\partial E / \partial X_{\text{C}}$} & \multicolumn{1}{c}{$\partial E / \partial X_{\text{H}_2}$} & \multicolumn{1}{c}{$|\sum\partial E / \partial X|$\tablenotemark[3]} & \multicolumn{1}{c}{$\partial E / \partial a$} & \multicolumn{1}{c}{$\partial E / \partial \theta$}  \\ \hline
No quad. derivatives & 8656 & $+0.0014519$ & $+0.0002503$ & $-0.0016847$ & $2\times10^{-5}$ & $+0.3906940$ & $-0.0025783$  \\
+ Surface integral & 8656 & & & & & $-0.0328248$ & \\
Quad. derivatives& 8656 & $+0.0013853$ & $+0.0002727$ & $-0.0016579$ & $< 10^{-10}$ & $-0.0328097$ & $-0.0099224$ \\
Finite differences & 8656 & $+0.0013854$ & $+0.0002725$ & $-0.0016579$ & $2\times10^{-9}$ & $-0.0328100$ &  $-0.0099231$ \\ \hline
No quad. derivatives & $25\times86$ & $+0.0014520$ & $+0.0001628$ & $-0.0016819$ & $7\times10^{-5}$ & $+0.3916527$ & $-0.0025377$  \\
+ Surface integral & $25\times86$ & & & & & $-0.0331890$ & \\
Quad. derivatives& $25\times86$ & $+0.0008567$ & $+0.0004540$ & $-0.0013107$ & $< 10^{-10}$ & $-0.0332414$ & $-0.0099663$ \\
Finite differences & $25\times86$ & $+0.0008567$ & $+0.0004540$ & $-0.0013107$ & $5\times10^{-8}$ & $-0.0332419$ & $-0.0099660$ \\ \hline
No quad. derivatives & $50\times194$ & $+0.0014287$ & $+0.0002459$ & $-0.0016757$ & $1\times10^{-6}$ & $+0.3907357$ & $-0.0026006$  \\
+ Surface integral & $50\times194$ & & & & & $-0.0327741$ & \\
Quad. derivatives& $50\times194$ & $+0.0016370$ & $+0.0000707$ & $-0.0017077$ & $1\times 10^{-10}$ & $-0.0326691$ & $-0.0099089$ \\  \hline
No quad. derivatives & $100\times302$ & $+0.0014530$ & $+0.0002447$ & $-0.0016828$ & $1\times10^{-5}$ & $+0.3907022$ & $-0.0025791$  \\
+ Surface integral & $100\times302$ & & & & & $-0.0328223$ & \\
Quad. derivatives& $100\times302$ & $+0.0014049$ & $+0.0002785$ & $-0.0016834$ & $1\times10^{-10}$ & $-0.0328142$ & $-0.0099251$ \\  \hline
No quad. derivatives & $200\times770$ & $+0.0014456$ & $+0.0002375$ & $-0.0016832$ & $9\times10^{-8}$ & $+0.3907088$ & $-0.0025805$  \\
+ Surface integral & $200\times770$ & & & & & $-0.0328205$ & \\
Quad. derivatives& $200\times770$ & $+0.0014464$ & $+0.0002370$ & $-0.0016835$ & $1 \times 10^{-10}$ & $-0.0328201$ & $-0.0099226$ \\  \hline
No quad. derivatives & $300\times974$ & $+0.0014457$ & $+0.0002375$ & $-0.0016832$ & $4\times10^{-9}$ & $+0.3907088$ & $-0.0025804$  \\
+ Surface integral & $300\times974$ & & & & & $-0.0328205$ & \\
Quad. derivatives& $300\times974$ & $+0.0014456$ & $+0.0002376$ & $-0.0016832$ & $< 10^{-10}$ & $-0.0328205$ & $-0.0099230$ \\
\end{tabular} 
\end{ruledtabular}
\tablenotetext[1]{See the corresponding footnote of Table \ref{6-31G}.} 
\tablenotetext[2]{See the corresponding footnote of Table \ref{6-31G}.} 
\tablenotetext[3]{See the corresponding footnote of Table \ref{6-31G}.} 
\end{table*}
%\end{squeezetable}

\begin{table}
\caption{The Cartesian coordinates (in bohr) of atoms, 
the lattice constant or ``rise'' $a$ (in bohr), and the helical angle or ``twist'' $\theta$ (in degrees) 
of the CH$_2$ repeat unit in helical polyethylene.\label{geom}}
\begin{ruledtabular}
\begin{tabular}{lrrr}
      & \multicolumn{1}{c}{$x$} & \multicolumn{1}{c}{$y$} & \multicolumn{1}{c}{$z$}  \\ \hline
H$_1$ & $0.1$ & $2.0$ & $1.0$  \\
C   & $0.0$ & $0.5$ & $0.0$  \\
H$_2$ & $0.0$ & $2.0$ & $-1.0$  \\
$a$ & $2.5$ \\
$\theta$ & $170.0$ \\
\end{tabular} 
\end{ruledtabular}
\end{table}

Some of the B3LYP/6-31G and B3LYP/cc-pVDZ energy gradients \cite{B3LYP} of helical polyethylene are given in Tables \ref{6-31G} and \ref{cc-pVDZ}, respectively. 
The structure of the helical repeat unit cell, which is a CH$_2$ unit, along with the lattice constant or ``rise'' $a$ and helical angle or ``twist'' $\theta$,
are given in Table \ref{geom}. 
Note that our B3LYP functional used the VWN5 local correlation functional. The Namur cutoff criterion was adopted 
with the short-range cutoff of $S=10$ and long-range cutoff of $L=30$ without a multipole-expansion correction \cite{Delhalle}. 
The number of wave vector ($k$) sampling points was 100 in the reciprocal unit cell. 
The integral screening threshold for the two-electron-integral evaluation \cite{Obara1986} was $1\times10^{-11}$\ a.u.  All calculations were performed with the {\sc polymer} programme \cite{polymer}.

The grid is defined only in the zeroth cell and  
consists of interlocking atom-centre grids \cite{Becke1988}, each of which is a product of the 
Gauss--Chebyshev radial grid \cite{Becke1988} and Lebedev angular grid \cite{Lebedev1975,Lebedev1976,Lebedev1977,Lebedev1992,Lebedev1995,Lebedev1999}. Their orientations are fixed in some user-defined coordinates (cf.\ Eq.\ (5) of Ref.\ \cite{Johnson1993})
rather than in the molecular axes (cf.\ Eq.\ (7) of Ref.\ \cite{JohnsonCPL1993}), thus making the energy and gradients  
vary slightly with the user's arbitrary choice of the coordinates since the Lebedev grids are not rotationally invariant. 
This orientation-dependence is minimized by
inclusion of the quadrature derivatives \cite{JohnsonCPL1993} and eventually eliminated in the limit of an infinitely dense grid. 
Conclusions of this article are unchanged by the lack of rotational invariance of the grids.

In-phase atomic gradients change noticeably upon inclusion of the quadrature derivatives for the small ($25\times86$) grid 
consisting of 25 radial points $\times$ 86 angular points per atom.
With increasing grid size, the differences between gradients with and without the quadrature derivatives taper below $2\times10^{-7}$\ a.u. 
Concomitantly, the deviations from the translational invariance (the absolute value of the sum of the $x$ atomic gradients) without 
the quadrature derivatives are on the order of $10^{-4}$\ a.u.\  or more for the $25\times86$ grid, but they decrease to $10^{-8}$ a.u.\ or less for the $300\times974$ grid.
With the quadrature derivatives, the deviations become on the order of the integral screening threshold regardless of grid size. 

Curiously, the atomic gradients without the quadrature derivatives seem consistently closer to the values in the limit of an infinitely dense grid. 
However, the atomic gradients with the quadrature derivatives are within $1 \times 10^{-7}$\ a.u.\ of the finite-difference benchmarks regardless of grid size, and are therefore more precise. 
%Both may have their utility in applications.
With a grid that is sufficiently large for energy calculations (the 8565 or $100\times302$ grid), the differences are negligible for atomic gradients, if not for 
frequency calculation \cite{JohnsonCPL1993,Shang2020,Malagoli2003}.

The lattice-constant gradient without the quadrature derivatives or surface-integral correction differs from the correct value by an order of magnitude and has the wrong sign
even for the gigantic $300\times974$ grid.
The addition of the surface integral brings it within $2\times10^{-4}$ a.u.\ of the correct value for the $25\times86$ grid and within less than $10^{-7}$ a.u.\ for the 
$300\times974$ grid. On the other hand, the deviations of the lattice-constant gradient with the quadrature derivatives from the finite-difference benchmarks 
are on the order of $10^{-7}$ a.u.\ for the 8656 or $25\times86$ grid (these slight deviations are due to the finite-difference approximation). 
The more precise agreement between the latter two methods indicates that the surface integral method is slightly approximate and the method including
the quadrature derivatives should always be trusted, although the error in the surface integral is negligible. 
Not including either the surface-integral correction or quadrature derivatives will lead to unacceptably large errors in the lattice-constant gradient, 
rendering the method useless.

The helical-angle gradient without the quadrature derivatives is wrong by the same order of magnitude as the gradient itself  regardless of grid size 
for this particular geometry.
The value with the quadrature derivatives agree with the finite-difference benchmark within a few $10^{-7}$ a.u.\ for the $8656$ or $25\times86$ grid, and
these small errors are ascribed to the finite-difference approximation. It is therefore crucial to include the quadrature derivatives in this gradient also.

\acknowledgments
This work was supported by the U.S. Department of Energy, Office of Science, Office of Basic Energy Sciences under Grant No.\ DE-SC0006028
and also by the centre for Scalable, Predictive methods for Excitation and Correlated phenomena (SPEC), which is funded by 
the U.S. Department of Energy, Office of Science, Office of Basic Energy Sciences, 
Chemical Sciences, Geosciences, and Biosciences Division, as a part of the Computational Chemical Sciences Program.
This research used resources of the National Energy Research Scientific Computing centre (NERSC), a U.S. Department of Energy Office of Science User Facility located at Lawrence Berkeley National Laboratory, operated under Contract No.\ DE-AC02-05CH11231 using NERSC award m3196 (2022).

\appendix

\section{Energy of a helical polymer\label{A}}

In atomic units, energy per helical repeat unit cell of an infinitely extended helix is written as
\begin{eqnarray}
E = E_{\text{NR}} + E_{\text{T}} + E_{\text{NA}} + E_{\text{J}} + c E_{\text{K}} + E_{\text{XC}} + E_{\text{MPE}}, \label{E}
\end{eqnarray}
which encompasses the spin-restricted HF theory ($c  = 1$ and $E_{\text{XC}}=0$) and a pure local or gradient-corrected DFT ($c = 0$ and $E_{\text{XC}} \neq 0$)
as well as a hybrid DFT such as B3LYP \cite{B3LYP} ($c \neq 0$ and $E_{\text{XC}} \neq 0$). Generalizing it to the spin-unrestricted case should be straightforward.

The first term is the nuclear-repulsion energy, which is written as
\begin{eqnarray}
E_{\text{NR}} &=& \frac{1}{2} \sum_{n=-L}^L  \sum_{I,J}^{\text{nuc.}}\frac{Z_IZ_J}{|\bm{R}_{I(0)} - \bm{R}_{J(n)}|},
\end{eqnarray}
where $I$ and $J$ run over all nuclei with nuclear charges of $Z_I$ and $Z_J$, and $n$
runs over cells ($-L \leq n \leq L$) with $L$ standing for the long-range lattice sum cutoff \cite{Delhalle}.
Summands with zero denominator are excluded from the summation.
$\bm{R}_{I(n)}$ is the position of the $I$th nucleus in the $n$th cell,
\begin{eqnarray}
\bm{R}_{I(n)} &=& \left(X_{I(n)}, Y_{I(n)}, Z_{I(n)}\right),
\end{eqnarray}
whose elements are defined by Eqs.\ (\ref{X})--(\ref{Z}). The $n$th cell is twisted anticlockwise by $n\theta$ relative to the zeroth cell. 

The second term is the kinetic energy per cell expressed as
\begin{eqnarray}
E_{\text{T}} &=&\sum_{n=-S}^S  \sum_{\mu,\nu}^{\text{cgs}} \tilde P_{\mu(0)\nu(n)} \tilde T_{\mu(0)\nu(n)} ,
\end{eqnarray}
where $\tilde P_{\mu(0)\nu(n)}$ is the density matrix element (defined at the end of this Appendix) for the $\mu$th contracted Gaussian-type orbital (cgs) in the zeroth cell and 
the $\nu$th cgs in the $n$th cell that translates and rotates with the helix (defined in the next paragraph).
$S$ is the short-range lattice sum cutoff \cite{Delhalle}.
$\tilde T_{\mu(0)\nu(n)}$ is the corresponding kinetic-energy integral, defined by
\begin{eqnarray}
\tilde T_{\mu(0)\nu(n)} =  \int d\bm{r}\, \tilde \chi^*_{\mu(0)}(\bm{r}) \left( -\frac{1}{2}\nabla^2\right) \tilde \chi_{\nu(n)}(\bm{r}) ,
\end{eqnarray}
where $\tilde \chi_{\nu(n)}(\bm{r})$ is the translating-rotating $\nu$th cgs in the $n$th cell. 
Generally, tilde indicates screw-axis-symmetry-adapted quantities throughout this article. 

Specifically, a screw-axis-symmetry-adapted cgs is given as a linear combination of the also screw-axis-symmetry-adapted 
primitive Gaussian-type orbital (pgs). It is therefore defined by
\begin{eqnarray}
\tilde \chi_{\kappa(n)}(\bm{r}) &=& \sum_\alpha^{\text{pgs}} c_{\kappa\alpha} 
\left(\tilde x_{(n)}\right)^{n_{\kappa x}} \left(\tilde y_{(n)}\right)^{n_{\kappa y}} \left(\tilde z_{(n)}\right)^{n_{\kappa z}}  
\nonumber\\&& \times\,
\exp\left(-\zeta_\alpha | \bm{r} - \bm{R}_{\kappa(n)} |^2 \right) ,
\label{cgs}
\end{eqnarray}
where $c_{\kappa\alpha}$ is the product of contraction coefficient and normalization factor of the $\alpha$th pgs,
$\zeta_\alpha$ is the exponent of the $\alpha$th pgs, and $\bm{R}_{\kappa(n)}$ is the coordinates of the nucleus in the $n$th cell on which the $\kappa$th cgs
and all of its constituent pgs are centred.

The coordinates $(\tilde x_{(n)},\tilde y_{(n)},\tilde z_{(n)})$ are the $n$th-cell-fixed Cartesian coordinates that rotate 
with the helix and are centred on $\bm{R}_{\kappa(n)} = ({X}_{\kappa(n)},{Y}_{\kappa(n)},{Z}_{\kappa(n)})$.
They are related to the space-fixed Cartesian coordinates $\bm{r} = (x,y,z)$ by
\begin{eqnarray}
\tilde x_{(n)} &=& x - X_{\kappa(n)} = x - na -  X_{\kappa(0)},  \label{tildex} \\
\tilde y_{(n)} &=& \left(y - Y_{\kappa(n)} \right) \cos n \theta + \left( z - Z_{\kappa(n)}\right) \sin n \theta 
\nonumber\\
&=& y \cos n \theta + z \sin n \theta - Y_{\kappa(0)}, \label{tildey} \\
\tilde z_{(n)} &=& - \left( y - Y_{\kappa(n)} \right) \sin n \theta + \left( z - Z_{\kappa(n)}\right) \cos n \theta 
\nonumber\\
&=& - y \sin n \theta + z \cos n \theta - Z_{\kappa(0)}. \label{tildez} 
\end{eqnarray}

Only with these screw-axis-symmetry-adapted (i.e., translating-rotating) basis functions does the Bloch theorem \cite{Kittel} hold, guaranteeing 
the translational invariance of matrices with tilde: 
\begin{eqnarray}
\tilde T_{\mu(m)\nu(n)}  &=& \tilde T_{\mu(0)\nu(n-m)},\\
\tilde P_{\mu(m)\nu(n)}  &=& \tilde P_{\mu(0)\nu(n-m)}.
\end{eqnarray}
However, various molecular integrals are more conveniently evaluated \cite{Obara1986} for cgs 
in the space-fixed Cartesian coordinates without tilde: %, which coincide with the rotating coordinates of the zeroth cell:
\begin{eqnarray}
 \chi_{\kappa(n)}(\bm{r}) &=& \sum_\alpha^{\text{pgs}} c_{\kappa\alpha} 
\left(x - X_{\kappa(n)}  \right)^{n_{\kappa x}} \left(y - Y_{\kappa(n)} \right)^{n_{\kappa y}} \left(z - Z_{\kappa(n)} \right)^{n_{\kappa z}} 
\nonumber\\&& \times\,
\exp\left(-\zeta_\alpha | \bm{r} - \bm{R}_{\kappa(n)} |^2 \right) .
\end{eqnarray}
The kinetic-energy-integral matrix in these non-symmetry-adapted cgs is defined by and computed as
\begin{eqnarray}
 T_{\mu(0)\nu(n)} &=&  \int d\bm{r}\,  \chi^*_{\mu(0)}(\bm{r}) \left( -\frac{1}{2}\nabla^2\right)  \chi_{\nu(n)}(\bm{r}),
\end{eqnarray}
using the standard algorithm such as the Obara--Saika method \cite{Obara1986}.
Generally, quantities without tilde designate those in the basis of the space-fixed Cartesian coordinates in this article. 

The screw-axis-symmetry-adapted kinetic-energy integrals are obtained from the non-symmetry-adapted ones by the
transformation,
\begin{eqnarray}
\tilde T_{\mu(0)\nu(n)} &=& \sum_{\nu^\prime}^{\text{cgs}}  R_{\nu\nu^\prime}(n) T_{\mu(0)\nu^\prime(n)}  ,
\end{eqnarray}
where $R_{\nu\nu^\prime}(n)$ is the rotation matrix for the $n$th cell. The latter is a unit matrix in which the 3-by-3 block in a 
normalized $p$ subshell (ordered $p_x$, $p_y$, $p_z$) is replaced by
\begin{eqnarray}
R_{\nu\nu^\prime}(n) = \left( 
\begin{array}{ccc}
1 & 0 & 0 \\
0 & c & s \\
0 & -s & c \\
\end{array}
\right),\label{rotation1}
\end{eqnarray}
with $c = \cos n\theta$ and $s=\sin n\theta$, the 6-by-6 block in 
a normalized Cartesian $d$ subshell (ordered $d_{xx}$, $d_{yy}$, $d_{zz}$, $d_{xy}$, $d_{yz}$, $d_{zx}$) is substituted by
\begin{eqnarray}
R_{\nu\nu^\prime}(n) = \left( 
\begin{array}{cccccc}
1 & 0 & 0 & 0 & 0 &0 \\
0 & c^2 & s^2 & 0 & {2cs}/{\sqrt{3}} & 0 \\
0 & s^2 & c^2 & 0 & -{2cs}/{\sqrt{3}} & 0\\
0 & 0 & 0 & c & 0 & s \\
0 & -\sqrt{3}cs & \sqrt{3}cs & 0 & c^2-s^2 & 0 \\
0 & 0 & 0 & -s & 0 & c \\
\end{array}
\right), \label{rotation2}
\end{eqnarray}
and the 10-by-10 block 
in a normalized Cartesian $f$ subshell (ordered $f_{xxx}$, $f_{yyy}$, $f_{zzz}$, $f_{xxy}$, $f_{yyz}$, $f_{zzx}$, $f_{xyy}$, $f_{yzz}$, $f_{zxx}$, $f_{xyz}$)
is swapped by
\begin{widetext}
\begin{eqnarray}
R_{\nu\nu^\prime}(n) = \left( 
\begin{array}{cccccccccc}
1 & 0 & 0 & 0 & 0 & 0 & 0 & 0 & 0 & 0 \\
0 & c^3 & s^3 & 0 & {3c^2s}/{\sqrt{5}} & 0 & 0 & {3cs^2}/{\sqrt{5}} & 0 & 0\\
0 & -s^3 & c^3 & 0 & {3cs^2}/{\sqrt{5}} & 0 & 0 & {-3c^2s}/{\sqrt{5}} & 0 & 0\\
0 & 0 & 0 & c & 0 & 0 & 0 & 0 & s & 0 \\
0 & -\sqrt{5}c^2s & \sqrt{5}cs^2 & 0 & c^3-2cs^2 & 0 & 0 & 2c^2s-s^3 & 0 & 0 \\
0 & 0 & 0 & 0 & 0 & c^2 & s^2 & 0 & 0 & -{2cs}/{\sqrt{3}} \\
0 & 0 & 0 & 0 & 0 & s^2 & c^2 & 0 & 0 & {2cs}/{\sqrt{3}} \\
0 & \sqrt{5}cs^2 & \sqrt{5}c^2s & 0 & s^3 -2c^2s & 0 & 0 & c^3-2cs^2 & 0 & 0 \\
0 & 0 & 0 & -s & 0 & 0 & 0 & 0 & c & 0 \\
0 & 0 & 0 & 0 & 0 & \sqrt{3}cs & -\sqrt{3}cs  & 0 & 0 & c^2-s^2 \\
\end{array}
\right). \label{rotation3}
\end{eqnarray}
\end{widetext}

The third term of Eq.\ (\ref{E}) is the nuclear-attraction energy given by
\begin{eqnarray}
E_{\text{NA}} &=&\sum_{n=-S}^S \sum_{\mu,\nu}^{\text{cgs}} \tilde P_{\mu(0)\nu(n)} \tilde N_{\mu(0)\nu(n)} ,
\end{eqnarray}
where the nuclear-attraction integrals over the translating-rotating basis functions are obtained by  the transformation,
\begin{eqnarray}
\tilde N_{\mu(0)\nu(n)} &=& \sum_{\nu^\prime}^{\text{cgs}}  R_{\nu\nu^\prime}(n) N_{\mu(0)\nu^\prime(n)}  .
\end{eqnarray}
The nuclear-attraction integrals are computed in the space-fixed Cartesian coordinates as
\begin{eqnarray}
N_{\mu(0)\nu(n)} = \sum_{m=-L}^L \sum_I^{\text{nuc.}}  \int d\bm{r}\,  \chi^*_{\mu(0)}(\bm{r})  \frac{-Z_I}{|\bm{r}-\bm{R}_{I(m)} |}  \chi_{\nu(n)}(\bm{r}).
\nonumber\\
\end{eqnarray}

The subsequent two terms in Eq.\ (\ref{E}) are the Coulomb and exchange energies given by
\begin{eqnarray}
E_{\text{J}} &=& \frac{1}{2}\sum_{n=-S}^S  \sum_{\mu,\nu}^{\text{cgs}} \tilde P_{\mu(0)\nu(n)} \tilde J_{\mu(0)\nu(n)} , \\
E_{\text{K}} &=& \frac{1}{2} \sum_{n=-S}^S \sum_{\mu,\nu}^{\text{cgs}} \tilde P_{\mu(0)\nu(n)} \tilde K_{\mu(0)\nu(n)} ,
\end{eqnarray}
where the so-called $J$ and $K$ integrals in the translating-rotating basis functions are obtained by the same transformation,
\begin{eqnarray}
\tilde J_{\mu(0)\nu(n)} &=& \sum_{\nu^\prime}^{\text{cgs}}  R_{\nu\nu^\prime}(n) J_{\mu(0)\nu^\prime(n)}  , \\
\tilde K_{\mu(0)\nu(n)} &=& \sum_{\nu^\prime}^{\text{cgs}}  R_{\nu\nu^\prime}(n)  K_{\mu(0)\nu^\prime(n)}  .
\end{eqnarray}
The $J$ and $K$ integrals in the space-fixed Cartesian coordinates are, in turn, built from two-electron integrals:
\begin{widetext}
\begin{eqnarray}
 J_{\mu(0)\nu(n)} &=&  \sum_{l=-L}^L \sum_{m=-S}^S \sum_{\kappa,\lambda}^{\text{cgs}}  P_{\kappa(l)\lambda(l+m)} 
\langle \mu(0)\nu(n) | \kappa(l)\lambda(l+m) \rangle , \\
 K_{\mu(0)\lambda(m)} &=& - \frac{1}{2} \sum_{l=-L}^L \sum_{n=-S}^S\sum_{\kappa,\lambda}^{\text{cgs}}  P_{\kappa(l)\nu(n)} 
\langle \mu(0)\nu(n) | \kappa(l)\lambda(m) \rangle ,
\end{eqnarray}
with 
\begin{eqnarray}
\langle \mu(0)\nu(n) | \kappa(l)\lambda(l+m) \rangle 
= \iint d\bm{r}_1d\bm{r}_2\,  \chi^*_{\mu(0)}(\bm{r}_1)  \chi_{\nu(n)}(\bm{r}_1)  \frac{1}{|\bm{r}_1-\bm{r}_2 |}  \chi^*_{\kappa(l)}(\bm{r}_2) \chi_{\lambda(l+m)}(\bm{r}_2),
\end{eqnarray}
\end{widetext}
in the Mulliken notation.
The density matrix in the space-fixed Cartesian coordinates entering the above equations is obtained by the back-transformation,
\begin{eqnarray}
 P_{\mu(l)\nu(m)} = \sum_{\mu^\prime,\nu^\prime}^{\text{cgs}} \tilde  P_{\mu^\prime(0)\nu^\prime(m-l)} R_{\mu^\prime\mu}(l) R_{\nu^\prime\nu}(m) , \label{P}
\end{eqnarray} 
 which is no longer expected to possess the translational invariance, i.e.,
\begin{eqnarray}
P_{\mu(l)\nu(m)} \neq  P_{\mu(0)\nu(l-m)}.
\end{eqnarray}

The penultimate term of Eq.\ (\ref{E}), $E_{\text{XC}}$, is the exchange-correlation (XC) energy evaluated by quadrature [see Eq.\ (\ref{exc_grid})].
%\begin{eqnarray}
%E_{\text{XC}} &=& \int_0^a dx \iint dydz \, f_{\text{XC}} (\rho_\alpha,\rho_\beta,\gamma_{\alpha\alpha},\gamma_{\alpha\beta},\gamma_{\beta\beta}; \bm{r}) \nonumber \\
%&\approx& \sum_{I}^{\text{nuc.}} \sum_{g(I)} w_I(\bm{r}_{g(I)})  f_{\text{XC}}(\bm{r}_{g(I)}).
%\end{eqnarray}
Integrand $f_{\text{XC}}$ has the same screw axis symmetry as the helix and so the XC energy per cell can be evaluated as 
its integral over the zeroth cell volume. In practice, an infinitely extended $f_{\text{XC}}$ is divided into cellwise contributions in a gradually overlapping manner by
Becke's fuzzy cell scheme.

The electron density and density gradient invariant are defined only within the zeroth cell (where the space-fixed and translating-rotating Cartesian coordinates 
coincide) and evaluated explicitly with the translating-rotating basis functions.
\begin{eqnarray}
\rho(\bm{r}) &=&\sum_{m,n=-S}^S \sum_{\mu,\nu}^{\text{cgs}}\tilde P_{\mu(0)\nu(n)} \tilde \chi^*_{\mu(m)}(\bm{r})\tilde  \chi_{\nu(m+n)}(\bm{r}), \label{rho} \\
\gamma(\bm{r}) &=& \nabla \rho(\bm{r}) \cdot \nabla \rho(\bm{r}) \nonumber\\ 
&=& \frac{\partial \rho(\bm{r})}{\partial x}\frac{\partial \rho(\bm{r})}{\partial x} + \frac{\partial \rho(\bm{r})}{\partial y}\frac{\partial \rho(\bm{r})}{\partial y}+ \frac{\partial \rho(\bm{r})}{\partial z}\frac{\partial \rho(\bm{r})}{\partial z},
\end{eqnarray}
with
%\begin{widetext}
\begin{eqnarray}
\frac{\partial \rho(\bm{r})}{\partial x} &=& \sum_{m,n=-S}^S \sum_{\mu,\nu}^{\text{cgs}} \tilde P_{\mu(0)\nu(n)}  
\tilde \chi^*_{\mu(m)}(\bm{r})  \frac{\partial \tilde\chi_{\nu(m+n)}(\bm{r})}{\partial \tilde x_{(m+n)}}
\nonumber \\&&
+\sum_{m,n=-S}^S \sum_{\mu,\nu}^{\text{cgs}} \tilde P_{\mu(0)\nu(n)}
\frac{\partial \tilde \chi^*_{\mu(m)}(\bm{r}) }{\partial \tilde x_{(m)}} \tilde\chi_{\nu(m+n)}(\bm{r}) ,
\end{eqnarray}
where Eq.\ (\ref{tildex}) was used, and
\begin{eqnarray}
\frac{\partial \rho(\bm{r})}{\partial y} &=& \sum_{m,n=-S}^S\sum_{\mu,\nu}^{\text{cgs}} \tilde P_{\mu(0)\nu(n)}  
\tilde \chi^*_{\mu(m)}(\bm{r})  \frac{\partial \tilde\chi_{\nu(m+n)}(\bm{r})}{\partial \tilde y_{(m+n)}} 
%\nonumber\\&&\times\, 
\cos(m+n)\theta
\nonumber \\&&
+ \sum_{m,n=-S}^S\sum_{\mu,\nu}^{\text{cgs}} \tilde P_{\mu(0)\nu(n)}  
\frac{\partial \tilde \chi^*_{\mu(m)}(\bm{r}) }{\partial \tilde y_{(m)}} \tilde\chi_{\nu(m+n)}(\bm{r}) \cos m\theta
\nonumber \\&& 
- \sum_{m,n=-S}^S\sum_{\mu,\nu}^{\text{cgs}}  \tilde P_{\mu(0)\nu(n)}  
\tilde \chi^*_{\mu(m)}(\bm{r})  \frac{\partial \tilde\chi_{\nu(m+n)}(\bm{r})}{\partial \tilde z_{(m+n)}} 
%\nonumber\\&&\times\, 
\sin(m+n)\theta 
\nonumber\\&&
- \sum_{m,n=-S}^S\sum_{\mu,\nu}^{\text{cgs}} \tilde P_{\mu(0)\nu(n)}  
\frac{\partial \tilde \chi^*_{\mu(m)}(\bm{r}) }{\partial \tilde z_{(m)}} \tilde\chi_{\nu(m+n)}(\bm{r}) \sin m\theta, \nonumber\\
\end{eqnarray}
as well as
\begin{eqnarray}
\frac{\partial \rho(\bm{r})}{\partial z} &=& \sum_{m,n=-S}^S\sum_{\mu,\nu}^{\text{cgs}} \tilde P_{\mu(0)\nu(n)}  
\tilde \chi^*_{\mu(m)}(\bm{r})  \frac{\partial \tilde\chi_{\nu(m+n)}(\bm{r})}{\partial \tilde y_{(m+n)}} \sin(m+n)\theta 
\nonumber\\&&
+ \sum_{m,n=-S}^S\sum_{\mu,\nu}^{\text{cgs}} \tilde P_{\mu(0)\nu(n)}  
\frac{\partial \tilde \chi^*_{\mu(m)}(\bm{r}) }{\partial \tilde y_{(m)}} \tilde\chi_{\nu(m+n)}(\bm{r}) \sin m\theta \nonumber\\
&& + \sum_{m,n=-S}^S\sum_{\mu,\nu}^{\text{cgs}} \tilde P_{\mu(0)\nu(n)}  
\tilde \chi^*_{\mu(m)}(\bm{r})  \frac{\partial \tilde\chi_{\nu(m+n)}(\bm{r})}{\partial \tilde z_{(m+n)}} 
%\nonumber\\&&\times\, 
\cos(m+n)\theta
 \nonumber\\&&
  + \sum_{m,n=-S}^S\sum_{\mu,\nu}^{\text{cgs}}  \tilde P_{\mu(0)\nu(n)}  
\frac{\partial \tilde \chi^*_{\mu(m)}(\bm{r}) }{\partial \tilde z_{(m)}} \tilde\chi_{\nu(m+n)}(\bm{r}) \cos m\theta. \nonumber\\
\end{eqnarray}
where Eqs.\ (\ref{tildey}) and (\ref{tildez}) were used. 

The XC integrals, necessary in the self-consistent field procedure (see below), are also evaluated by quadrature.
\begin{eqnarray}
\tilde X_{\mu(0)\nu(n)} &=&  \int d\bm{r}\, \tilde  \chi^*_{\mu(0)}(\bm{r}) v_{\text{XC}}(\bm{r})\tilde \chi_{\nu(n)}(\bm{r}) 
\nonumber \\&& 
+ 2 \int d\bm{r}\,  \tilde \chi^*_{\mu(0)}(\bm{r})  u_{\text{XC}}(\bm{r}) \nabla \rho (\bm{r}) \cdot \nabla \tilde \chi_{\nu(n)}(\bm{r}) 
\nonumber\\&& 
+ 2 \int d\bm{r}\,  \nabla \tilde \chi^*_{\mu(0)}(\bm{r}) \cdot \nabla \rho(\bm{r}) {u}_{\text{XC}}(\bm{r}) \,\tilde \chi_{\nu(n)}(\bm{r}) \label{XCint1} \\
&\approx& \sum_{m=-S}^S\sum_{I}^{\text{nuc.}} \sum_{g(I)} w_I(\bm{r}_{g(I)}) \tilde  \chi^*_{\mu(m)}(\bm{r}_{g(I)}) 
\nonumber\\ && \times\,v_{\text{XC}}(\bm{r}_{g(I)})\tilde \chi_{\nu(m+n)}(\bm{r}_{g(I)}) \nonumber\\
&&+ 2 \sum_{m=-S}^S\sum_{I}^{\text{nuc.}} \sum_{g(I)} w_i(\bm{r}_{g(I)}) 
\tilde  \chi^*_{\mu(m)}(\bm{r}_{g(I)}) {u}_{\text{XC}}(\bm{r}_{g(I)}) 
\nonumber\\ && \times\,\nabla \rho (\bm{r}_{g(I)}) \cdot \nabla \tilde \chi_{\nu(m+n)}(\bm{r}_{g(I)}) \nonumber\\
&&+ 2 \sum_{m=-S}^S\sum_{I}^{\text{nuc.}} \sum_{g(I)} w_i(\bm{r}_{g(I)}) 
\nabla\tilde  \chi^*_{\mu(m)}(\bm{r}_{g(I)}) \cdot  \nabla \rho (\bm{r}_{g(I)}) 
\nonumber\\ && \times\,u_{\text{XC}}(\bm{r}_{g(I)}) \tilde \chi_{\nu(m+n)}(\bm{r}_{g(I)}), \label{XCint2}
\end{eqnarray}
%\end{widetext}
where the XC potentials are defined only within the zeroth cell as
\begin{eqnarray}
v_{\text{XC}}(\bm{r}) &=& \frac{\partial f_{\text{XC}}(\bm{r})}{\partial \rho_\alpha(\bm{r})}, \\
u_{\text{XC}}(\bm{r}) &=& 
\frac{1}{2} \frac{\partial f_{\text{XC}}(\bm{r})}{\partial \gamma_{\alpha\alpha}(\bm{r})} + \frac{1}{4} \frac{\partial f_{\text{XC}}(\bm{r})}{\partial \gamma_{\alpha\beta}(\bm{r})}. \end{eqnarray}
The derivatives of the translating-rotating basis functions with respect to the space-fixed Cartesian coordinates are evaluated as
\begin{eqnarray}
\frac{\partial \tilde \chi_{\nu(n)}(\bm{r})}{\partial x} &=& \frac{\partial \tilde\chi_{\nu(n)}(\bm{r})}{\partial \tilde x_{(n)}} , \label{chi_x} \\
\frac{\partial \tilde \chi_{\nu(n)}(\bm{r})}{\partial y} &=& \frac{\partial \tilde\chi_{\nu(n)}(\bm{r})}{\partial \tilde y_{(n)}} \cos n\theta
- \frac{\partial \tilde\chi_{\nu(n)}(\bm{r})}{\partial \tilde z_{(n)}} \sin n\theta ,\label{chi_y} \\
\frac{\partial \tilde \chi_{\nu(n)}(\bm{r})}{\partial z} &=& \frac{\partial \tilde\chi_{\nu(n)}(\bm{r})}{\partial \tilde y_{(n)}} \sin n\theta
+ \frac{\partial \tilde\chi_{\nu(n)}(\bm{r})}{\partial \tilde z_{(n)}} \cos n\theta .\label{chi_z}
\end{eqnarray}
The derivatives in the right-hand sides are computed on each grid point explicitly. 

In Eqs.\ (\ref{rho})--(\ref{XCint2}), two translating-rotating basis functions are generally centred in different cells, while $v_{\text{XC}}$ is infinitely extended and periodic. 
Therefore, in principle, one needs to integrate the respective functions in all cells, using grid that is translated and rotated
 along the helix. An equivalent integration is achieved by defining grid only within the zeroth cell and translate and rotate the two basis functions 
 along the helix. We have chosen the latter method and the above equations reflect this choice.

The last term, $E_{\text{MPE}}$, stands for the multipole-expansion correction \cite{Delhalle}. In the leading order, it is given by
\begin{eqnarray}
E_{\text{MPE}} = \sum_{n=L+1}^{\infty} \frac{1}{(na)^3} \left (-2 \mu_x^2 + \mu_y^2 \cos n\theta+ \mu_z^2 \cos n\theta \right ), \nonumber\\
\end{eqnarray}
where $(\mu_x,\mu_y,\mu_z)$ is related to the zeroth-cell dipole moment, and $\infty$ is mimicked by a large integer (10,000 in our implementation). 
Note that the charge-quadruple interactions, which also decay as $r^{-3}$ and were recommended for inclusion by Delhalle {\it et al.}\ \cite{Delhalle},
are excluded in this study, so that the energy is variational and consistent with the analytical gradients (see Appendix \ref{B}).
The corresponding correction to the Kohn--Sham Hamiltonian (or Fock) matrix elements are
\begin{eqnarray}
\tilde M_{\mu(0)\nu(n)} &=& \sum_{\nu^\prime}^{\text{cgs}}  R_{\nu\nu^\prime}(n) M_{\mu(0)\nu^\prime(n)} 
\end{eqnarray}
with
\begin{eqnarray}
 M_{\mu(0)\nu(n)} 
&=&\sum_{m=L+1}^{\infty} \frac{1}{(ma)^3} 
\left \{ %S_{\mu(0)\nu(n)} \, q_{3x^2-r^2} 
-4 x_{\mu(0)\nu(n)} \, \mu_x 
+ 2 y_{\mu(0)\nu(n)} \, \mu_y \cos m \theta 
\right. \nonumber\\&& \left. 
+ 2 z_{\mu(0)\nu(n)} \,  \mu_z \cos m \theta  \right \} .
\end{eqnarray}
Integrals $x_{\mu(0)\nu(n)}$, $y_{\mu(0)\nu(n)}$, and $z_{\mu(0)\nu(n)}$ are the moment integrals of the space-fixed Cartesian coordinates
in the nonrotating basis functions. For instance, a $y$ moment integral is defined by and evaluated as 
\begin{eqnarray}
 y_{\mu(0)\nu(n)} &=&  \int d\bm{r}\,  \chi^*_{\mu(0)}(\bm{r}) y\, \chi_{\nu(n)}(\bm{r}),
\end{eqnarray}
which, in turn, gives the $y$ ``dipole moment'' for the zeroth cell as
\begin{eqnarray}
\mu_y = \sum_{n=-S}^S  \sum_{\mu,\nu}^{\text{cgs}} P_{\mu(0)\nu(n)} y_{\mu(0)\nu(n)} -\sum_I^{\text{nuc.}}  Z_I y_I ,
\end{eqnarray}
where the density matrix in the space-fixed Cartesian coordinates is given by Eq.\ (\ref{P}). 

The Kohn--Sham Hamiltonian (or Fock) and overlap matrices in the translating-rotating basis functions are defined by
\begin{eqnarray}
\tilde F_{\mu(0)\nu(n)} &=& \tilde T_{\mu(0)\nu(n)} + \tilde N_{\mu(0)\nu(n)} + \tilde J_{\mu(0)\nu(n)} + c\tilde  K_{\mu(0)\nu(n)} 
\nonumber\\ &&
+\tilde  X_{\mu(0)\nu(n)}+\tilde  M_{\mu(0)\nu(n)},  
\end{eqnarray}
and
\begin{eqnarray}
\tilde S_{\mu(0)\nu(n)} &=& \sum_{\nu^\prime}^{\text{cgs}}  R_{\nu\nu^\prime}(n) S_{\mu(0)\nu^\prime(n)}  ,
\end{eqnarray}
with
\begin{eqnarray}
 S_{\mu(0)\nu(n)} &=&  \int d\bm{r}\,  \chi^*_{\mu(0)}(\bm{r})  \chi_{\nu(n)}(\bm{r}).
\end{eqnarray}
The dynamical Fock and overlap matrices are then constructed as
\begin{eqnarray}
\tilde F_{\mu\nu}(k) &=& \sum_{n=-S}^S \tilde F_{\mu(0)\nu(n)} \exp(inka), \\
\tilde S_{\mu\nu}(k) &=& \sum_{n=-S}^S \tilde S_{\mu(0)\nu(n)} \exp(inka),
\end{eqnarray}
where $k$ is the wave vector ($-\pi/a \leq k < \pi/a$).
At this point, the dynamical matrices in the spherical $d$ and $f$ translating-rotating basis functions (if applicable) can be obtained by the transformations,
\begin{eqnarray}
\bar{\tilde F}_{\mu\nu}(k) &=& \sum_{\mu^\prime,\nu^\prime}^{\text{cgs}} Q_{\mu\mu^\prime}Q_{\nu\nu^\prime}\tilde F_{\mu^\prime\nu^\prime}(k), \\
\bar{\tilde S}_{\mu\nu}(k) &=& \sum_{\mu^\prime,\nu^\prime}^{\text{cgs}} Q_{\mu\mu^\prime}Q_{\nu\nu^\prime}\tilde S_{\mu^\prime\nu^\prime}(k).
\end{eqnarray}
The transformation matrix is a unit matrix in which the 6-by-6 block in a normalized Cartesian $d$ subshell (in the same order as before)
is replaced by 
\begin{eqnarray}
Q_{\mu^\prime\mu}(n) = \left( 
\begin{array}{cccccc}
{\sqrt{3}}/{2} & -{\sqrt{3}}/{2} & 0 & 0 & 0 & 0 \\
-{1}/{2} & -{1}/{2} & 1 & 0 & 0 &0 \\
0 & 0 & 0 & 1 & 0 & 0 \\
0 & 0 & 0 & 0 & 1 & 0 \\
0 & 0 & 0 & 0 & 0 & 1 \\
0 & 0 & 0 & 0 & 0 & 0 \\
\end{array}
\right),
\end{eqnarray}
or the 10-by-10 block in a normalized Cartesian $f$ subshell (in the same order as before) is swapped by
\begin{widetext}
\begin{eqnarray}
Q_{\mu^\prime\mu}(n) = \left( 
\begin{array}{cccccccccc}
0 & 0 & 1 & 0 & -{3\sqrt{5}}/{10} & 0 & 0 & 0 & -{3\sqrt{5}}/{10} & 0 \\
-{\sqrt{6}}/{4} & 0 & 0 & 0 & 0 &  {\sqrt{30}}/{5} & -{\sqrt{30}}/{20} & 0 & 0 & 0 \\
0 & -{\sqrt{6}}/{4} & 0 & -{\sqrt{30}}/{20} & 0 & 0 & 0 & {\sqrt{30}}/{5} & 0 & 0 \\
0 & 0 & 0 & 0 & -{\sqrt{3}}/{2} & 0 & 0 & 0 & {\sqrt{3}}/{2} & 0 \\
0 & 0 & 0 & 0 & 0 & 0 & 0 & 0 & 0 & 1 \\
{\sqrt{10}}/{4} & 0 & 0 & 0 & 0 & 0 & -{3\sqrt{2}}/{4} & 0 & 0 & 0 \\
0 & -{\sqrt{10}}/{4} & 0 & {3\sqrt{2}}/{4} & 0 & 0 & 0 & 0 & 0 & 0 \\
0 & 0 & 0 & 0 & 0 & 0 & 0 & 0 & 0 & 0 \\
0 & 0 & 0 & 0 & 0 & 0 & 0 & 0 & 0 & 0 \\
0 & 0 & 0 & 0 & 0 & 0 & 0 & 0 & 0 & 0 \\
\end{array}
\right).
\end{eqnarray}
\end{widetext}

The generalized eigenvalue equation of the form,
\begin{eqnarray}
\sum_{\nu}^{\text{cgs}} \bar{\tilde F}_{\mu\nu}(k) \bar{\tilde C}_{\nu m}(k) = \sum_{\nu}^{\text{cgs}} \bar{\tilde S}_{\mu\nu}(k) \bar{\tilde C}_{\nu m}(k) \epsilon_{m}(k),
\end{eqnarray}
is solved with the L\"{o}wdin orthogonalization method \cite{szabo} 
(at each of the $K$ evenly spaced wave vector $k$ in the reciprocal unit cell) for the energy bands $\epsilon_{m}(k)$ and 
Bloch orbital coefficients $\bar{\tilde C}_{\nu m}(k)$ in the spherical $d$ and $f$ translating-rotating basis functions. 
The latter are transformed to those in the Cartesian $d$ and $f$ translating-rotating basis functions by
\begin{eqnarray}
\tilde C_{\nu m}(k) = \sum_{\nu^\prime}^{\text{cgs}} \bar{\tilde C}_{\nu^\prime m}(k) Q_{\nu^\prime\nu}.
\end{eqnarray}
The density matrix in the Cartesian $d$ and $f$ translating-rotating basis functions is then given by
\begin{eqnarray}
\tilde P_{\mu(0)\nu(n)} = \frac{2}{K} \sum_{m}^{\text{cgs}}\sum_{k}^{K} f_{m}(k) \tilde C^*_{\mu m}(k) \tilde C_{\nu m}(k) \exp(inka), \nonumber\\
\end{eqnarray}
with $f_{m}(k)$ being the Fermi--Dirac distribution function,
\begin{eqnarray}
f_{m}(k) = \frac{1}{\exp[\beta\{\epsilon_{m}(k)-\epsilon_{\text{F}}\}]+1},
\end{eqnarray}
where $\beta = (k_\text{B}T)^{-1}$ 
and the Fermi energy $\epsilon_{\text{F}}$  is determined by the electro-neutrality condition:
\begin{eqnarray}
\sum_I Z_I = \sum_{n=-S}^{S}\sum_{\mu,\nu} \tilde P_{\mu(0)\nu(n)}  \tilde S_{\mu(0)\nu(n)} .
\end{eqnarray}

\section{Energy gradients of a helical polymer with respect to the helical angle \label{B}}

We shall consider the gradient of the total energy per cell with respect to the helical angle $\theta$ as the most complex example:
\begin{eqnarray}
\frac{\partial E}{\partial \theta} = \frac{\partial E}{\partial \theta}^{(h)} +\frac{\partial E}{\partial \theta}^{(b)} + \frac{\partial E}{\partial \theta}^{(d)} +  \frac{\partial E}{\partial \theta}^{(q)}.
\end{eqnarray}
Gradients with respect to
the in-phase atomic coordinates $X_{I(0)}$, $Y_{I(0)}$, $Z_{I(0)}$, or to the lattice constant $a$ can be inferred from it.
The first term in the right-hand side of the above equation is the Hellmann--Feynman force, the second term has to do with
the derivatives of basis functions (but not through grid coordinates), the third term 
 the density matrix derivatives, and the fourth term the quadrature derivatives.  
 
The Hellmann--Feynman force consists of two terms:
\begin{eqnarray}
\frac{\partial E}{\partial \theta}^{(h)} = \frac{\partial E_{\text{NR}}}{\partial \theta} + \frac{\partial E_{\text{NA}}}{\partial \theta}^{(h)},
\end{eqnarray}
each of which is easily formulated and will not be repeated here.
The density matrix derivatives are consolidated into the so-called Pulay force by virtue of the variational nature 
of the HF or DFT energy. It is given by
\begin{eqnarray}
\frac{\partial E}{\partial \theta}^{(d)} = - \sum_{n=-S}^S \sum_{\mu,\nu}^{\text{cgs}} \tilde W_{\mu(0)\nu(n)} \frac{\partial \tilde S_{\mu(0)\nu(n)}}{\partial \theta}, \label{Pulay}
\end{eqnarray}
with the energy-weighted density matrix, 
\begin{eqnarray}
\tilde W_{\mu(0)\nu(n)} = \frac{2}{K} \sum_{m}^{\text{cgs}}\sum_{k}^{K} f_{m}(k) \epsilon_{m}(k) \tilde C^*_{\mu m}(k) \tilde C_{\nu m}(k) \exp(inka), \nonumber\\
\end{eqnarray}
The quadrature derivatives exist only in the XC energy gradient and are extensively discussed in the main text.
\begin{eqnarray}
\frac{\partial E}{\partial \theta}^{(q)} = \frac{\partial E_{\text{XC}} }{\partial \theta}^{(q)} .
\end{eqnarray}

Therefore, in the remainder of this Appendix, we shall focus on the basis function derivatives:
\begin{eqnarray}
\frac{\partial E}{\partial \theta}^{(b)}&=& \frac{\partial E_{\text{T}}}{\partial \theta}^{(b)} + \frac{\partial E_{\text{NA}}}{\partial \theta}^{(b)} + \frac{\partial E_{\text{J}} }{\partial \theta}^{(b)}+ c \frac{\partial E_{\text{K}} }{\partial \theta}^{(b)}
\nonumber\\&&
+ \frac{\partial E_{\text{XC}} }{\partial \theta}^{(b)} + \frac{\partial E_{\text{MPE}} }{\partial \theta}^{(b)}. 
\end{eqnarray}

The basis function derivative of the kinetic energy with respect to $\theta$ is given by
\begin{eqnarray}
\frac{\partial E_{\text{T}}}{\partial \theta}^{(b)} &=&\sum_{n=-S}^S \sum_{\mu,\nu}^{\text{cgs}} \tilde P_{\mu(0)\nu(n)} \frac{\partial \tilde T_{\mu(0)\nu(n)} }{\partial \theta} \\
&=&\sum_{n=-S}^S \sum_{\mu,\nu}^{\text{cgs}} \sum_{\nu^\prime}^{\text{cgs}}  \tilde P_{\mu(0)\nu(n)} 
\nonumber\\&&\times\,
\left\{\frac{\partial R_{\nu\nu^\prime}(n) }{\partial \theta}  T_{\mu(0)\nu^\prime(n)}  + 
 R_{\nu\nu^\prime}(n) \frac{\partial   T_{\mu(0)\nu^\prime(n)}  }{\partial \theta} \right\} \\
 &=& \sum_{n=-S}^S\sum_{\mu,\nu}^{\text{cgs}} \sum_{\nu^\prime}^{\text{cgs}}  \tilde P_{\mu(0)\nu(n)}
\frac{\partial R_{\nu\nu^\prime}(n) }{\partial \theta}  T_{\mu(0)\nu^\prime(n)}  
\nonumber\\&&
+  \sum_{n=-S}^S \sum_{\mu,\nu}^{\text{cgs}}
  P_{\mu(0)\nu(n)} \frac{\partial   T_{\mu(0)\nu(n)}  }{\partial \theta}. \label{Tderiv}
\end{eqnarray}
In the last expression, the first term accounts for the basis function rotation, while the second for the basis function centre shift.
Equation (\ref{wrong}) is wrong for it neglects the first term.
The derivative of the kinetic-energy integral in the space-fixed Cartesian coordinates with respect to $\theta$ is computed as
\begin{eqnarray}
\frac{\partial   T_{\mu(0)\nu(n)}  }{\partial \theta} &=& 
\int d\bm{r}\,  \chi^*_{\mu(0)}(\bm{r}) \left( -\frac{1}{2}\nabla^2\right)  \frac{\partial  \chi_{\nu(n)}(\bm{r})}{\partial \theta} \\
&=& \int d\bm{r}\,  \chi^*_{\mu(0)}(\bm{r}) \left( -\frac{1}{2}\nabla^2\right) 
\nonumber\\&&\times\,
\left\{  \frac{\partial Y_{\nu(n)}}{\partial \theta}\frac{\partial  \chi_{\nu(n)}(\bm{r})}{\partial Y_{\nu(n)}} 
+  \frac{\partial Z_{\nu(n)}}{\partial \theta}\frac{\partial  \chi_{\nu(n)}(\bm{r})}{\partial Z_{\nu(n)}}  \right\} \nonumber\\ \\
&=& (-n Y_{\nu(0)}\sin n\theta -n Z_{\nu(0)}\cos n\theta) 
\nonumber\\&&\times\, \int d\bm{r}\,  \chi^*_{\mu(0)}(\bm{r}) \left( -\frac{1}{2}\nabla^2\right)  
 \left(-\frac{\partial  \chi_{\nu(n)}(\bm{r})}{\partial y} \right)  \nonumber \\
 && +  (n Y_{\nu(0)}\cos n\theta - n Z_{\nu(0)}\sin n\theta) 
 \nonumber\\&&\times\, \int d\bm{r}\,  \chi^*_{\mu(0)}(\bm{r}) \left( -\frac{1}{2}\nabla^2\right) 
 \left(-\frac{\partial  \chi_{\nu(n)}(\bm{r})}{\partial z} \right) ,  
\end{eqnarray}
where we used the fact that the basis functions in the space-fixed Cartesian coordinates depend on $\theta$ only through their centres 
and that $\partial / \partial Y = -\partial / \partial y$, etc. The molecular integral derivatives appearing in the last expression are evaluated by
the standard method \cite{Obara1986}.
The corresponding derivatives of $\tilde S_{\mu(0)\nu(n)}$ and $\tilde N_{\mu(0)\nu(n)}$ are obtained 
in the same way.

The basis function derivative of the Coulomb energy is given by
%\begin{widetext}
\begin{eqnarray}
\frac{\partial E_{\text{J}}}{\partial \theta}^{(b)} &=& 
\frac{1}{2}\sum_{l=-L}^L \sum_{m,n=-S}^S   \sum_{\mu,\nu,\kappa,\lambda}^{\text{cgs}} 
\nonumber\\&&\times\,
\frac{\partial}{\partial \theta}^{(b)} \left\{   P_{\mu(0)\nu(n)}   P_{\kappa(l)\lambda(l+m)} 
\langle \mu(0)\nu(n) | \kappa(l)\lambda(l+m) )\rangle \right\}\nonumber\\ \\
&=& \frac{1}{2} \sum_{l=-L}^L \sum_{m,n=-S}^S   \sum_{\mu,\nu,\kappa,\lambda}^{\text{cgs}} 
\nonumber\\&&\times\,
 \left\{  \frac{\partial  P_{\mu(0)\nu(n)}}{\partial \theta}^{(b)}    P_{\kappa(l)\lambda(l+m)} 
\langle \mu(0)\nu(n) | \kappa(l)\lambda(l+m) \rangle \right\} \nonumber \\
&& + \frac{1}{2} \sum_{l=-L}^L \sum_{m,n=-S}^S   \sum_{\mu,\nu,\kappa,\lambda}^{\text{cgs}} 
\nonumber\\&&\times\,
 \left\{  P_{\mu(0)\nu(n)} \frac{\partial  P_{\kappa(l)\lambda(l+m)} }{\partial \theta}^{(b)}   
\langle \mu(0)\nu(n) | \kappa(l)\lambda(l+m) \rangle \right\} \nonumber\\
&& + \frac{1}{2} \sum_{l=-L}^L \sum_{m,n=-S}^S   \sum_{\mu,\nu,\kappa,\lambda}^{\text{cgs}} 
\nonumber\\&&\times\,
 \left\{  P_{\mu(0)\nu(n)}  P_{\kappa(l)\lambda(l+m)} 
\frac{\partial  }{\partial \theta} 
\langle \mu(0)\nu(n) | \kappa(l)\lambda(l+m) \rangle \right\}. \nonumber\\
\end{eqnarray}
%\end{widetext}
The basis function derivative explicitly neglects the variation of density matrix elements. Hence, what is meant by the basis function derivative
of the density matrix in the space-fixed Cartesian coordinates is as follows:
\begin{eqnarray}
\frac{\partial  P_{\kappa(l)\lambda(l+m)}}{\partial \theta}^{(b)}  
&=& \sum_{\kappa^\prime,\lambda^\prime}^{\text{cgs}}  \tilde P_{\kappa^\prime(0)\lambda^\prime(m)} 
\frac{\partial R_{\kappa^\prime\kappa}(l)  }{\partial \theta} R_{\lambda^\prime\lambda}(l+m)  
\nonumber \\&& 
+ \sum_{\kappa^\prime,\lambda^\prime}^{\text{cgs}}  \tilde P_{\kappa^\prime(0)\lambda^\prime(m)} 
R_{\kappa^\prime\kappa}(l)  \frac{\partial R_{\lambda^\prime\lambda}(l+m) }{\partial \theta} , \nonumber\\ \label{Pder}
\end{eqnarray}
which includes only the effects of basis function rotations on the density matrix, but not the change in the density matrix ($\tilde P$) itself. 
The derivative of a two-electron integral with respect to $\theta$ is given by
%\begin{widetext}
\begin{eqnarray}
&&  \frac{\partial  }{\partial \theta}   
( \mu(0)\nu(n) | \kappa(l)\lambda(l+m) ) 
 \nonumber\\
&&= (-n Y_{\nu(0)}\sin n\theta -n Z_{\nu(0)}\cos n\theta) \langle \mu(0)\left(-\frac{\partial \nu(n)}{\partial y}\right) | \kappa(l)\lambda(l+m) \rangle 
\nonumber\\
&& +  (n Y_{\nu(0)}\cos n\theta - n Z_{\nu(0)}\sin n\theta) \langle \mu(0)\left(-\frac{\partial \nu(n)}{\partial z}\right) | \kappa(l)\lambda(l+m) \rangle 
\nonumber\\
&& + (-l Y_{\kappa(0)}\sin l\theta -l Z_{\kappa(0)}\cos l\theta) \langle \mu(0)\nu(n) | \left(-\frac{\partial  \kappa(l)}{\partial y}\right)\lambda(l+m) \rangle 
\nonumber\\
&& +  (l Y_{\kappa(0)}\cos l\theta - l Z_{\kappa(0)}\sin l\theta) \langle \mu(0)\nu(n) | \left(-\frac{\partial \kappa(l)}{\partial z}\right)\lambda(l+m) \rangle 
\nonumber\\
&& + \left\{ -(l+m) Y_{\lambda(0)}\sin (l+m)\theta -(l+m) Z_{\lambda(0)}\cos (l+m)\theta \right\}
\nonumber\\ &&\times\,
 \langle \mu(0)\nu(n) |  \kappa(l)\left(-\frac{\partial \lambda(l+m)}{\partial y}\right) \rangle 
\nonumber\\
&& +  \left\{ (l+m) Y_{\lambda(0)}\cos (l+m)\theta - (l+m) Z_{\lambda(0)}\sin (l+m)\theta \right\} 
\nonumber\\ &&\times\,
\langle \mu(0)\tilde\nu(n) |  \kappa(l)\left(-\frac{\partial\lambda(l+m) }{\partial z}\right)\rangle .
\end{eqnarray}
%\end{widetext}
The corresponding derivative of the exchange energy can be obtained in the same way.

The basis function derivative of the XC energy reads
\begin{eqnarray}
{\frac{\partial E_{\text{XC}}}{\partial \theta}}^{(b)} &=& {\frac{\partial }{\partial \theta}}^{(b)} \sum_{I}^{\text{nuc.}}\sum_{g(I)} w_I(\bm{r}_{g(I)}) f_{\text{XC}}(\bm{r}_{g(I)}) \\
&=&  %\sum_{I}^{\text{nuc.}}\sum_{g(I)} {\frac{\partial w_I(\bm{r}_{g(I)}) }{\partial \theta}}^{(b)}  f_{\text{XC}}(\bm{r}_{g(I)}) \nonumber \\ && +  
\sum_{I}^{\text{nuc.}}\sum_{g(I)} w_I(\bm{r}_{g(I)}) {\frac{\partial f_{\text{XC}}(\bm{r}_{g(I)})}{\partial \theta}}^{(b)} \\
&=&  %\sum_{I}^{\text{nuc.}}\sum_{g(I)} {\frac{\partial w_I(\bm{r}_{g(I)}) }{\partial \theta}}^{(b)}  f_{\text{XC}}(\bm{r}_{g(I)}) \nonumber \\&& + 
 \sum_{I}^{\text{nuc.}}\sum_{g(I)} w_I(\bm{r}_{g(I)})   v_{\text{XC}}(\bm{r}_{g(I)}) { \frac{\partial \rho(\bm{r}_{g(I)})}{\partial \theta} }^{(b)} \nonumber \\
&& +  \sum_{I}^{\text{nuc.}}\sum_{g(I)} w_I(\bm{r}_{g(I)})   u_{\text{XC}}(\bm{r}_{g(I)}) { \frac{\partial \gamma(\bm{r}_{g(I)})}{\partial \theta} }^{(b)} .
\end{eqnarray}
The derivative of the density is given by
\begin{eqnarray}
{ \frac{\partial \rho(\bm{r})}{\partial \theta} }^{(b)} 
&=&  \sum_{m,n=-S}^S  \sum_{\mu,\nu}^{\text{cgs}}
\tilde P_{\mu(0)\nu(n)} \tilde \chi^*_{\mu(m)}(\bm{r}) \frac{\partial \tilde \chi_{\nu(m+n)}(\bm{r}) }{\partial \theta}\nonumber \\
&& + \sum_{m,n=-S}^S  \sum_{\mu,\nu}^{\text{cgs}} 
\tilde P_{\mu(0)\nu(n)} \frac{\partial  \tilde \chi^*_{\mu(m)}(\bm{r}) }{\partial \theta} \tilde \chi_{\nu(m+n)}(\bm{r})  , \nonumber\\
\end{eqnarray}
with the derivatives of the translating-rotating basis functions computed as
\begin{eqnarray}
\frac{\partial \tilde \chi_{\mu(n)}(\bm{r}) }{\partial \theta} 
&=& n\left( - y \sin n\theta +  z \cos n\theta  \right) \frac{\partial \tilde \chi_{\mu(n)}(\bm{r}) }{\partial \tilde y_{(n)}} \nonumber \\
&& + n\left( - y \cos n\theta -  z \sin n\theta  \right) \frac{\partial \tilde \chi_{\mu(n)}(\bm{r}) }{\partial \tilde z_{(n)}} . \label{chi_theta}
\end{eqnarray}
The derivative of the density gradient invariant is
\begin{eqnarray}
{ \frac{\partial \gamma(\bm{r})}{\partial \theta} }^{(b)} 
&=& 2 \frac{\partial \rho(\bm{r})}{\partial x} { \frac{\partial^2 \rho(\bm{r})}{\partial \theta \partial x} }^{(b)} 
%\nonumber \\&&
+ 2 \frac{\partial \rho(\bm{r})}{\partial y} { \frac{\partial^2 \rho(\bm{r})}{\partial \theta \partial y} }^{(b)} 
\nonumber\\&&
+ 2 \frac{\partial \rho(\bm{r})}{\partial z} { \frac{\partial^2 \rho(\bm{r})}{\partial \theta \partial z} }^{(b)} , 
\end{eqnarray}
with 
\begin{eqnarray}
{ \frac{\partial^2 \rho(\bm{r})}{\partial \theta \partial x} }^{(b)} 
&=&  \sum_{m,n=-S}^S \sum_{\mu,\nu}^{\text{cgs}}
\tilde P_{\mu(0)\nu(n)} \tilde \chi^*_{\mu(m)}(\bm{r}) \frac{\partial^2 \tilde \chi_{\nu(m+n)}(\bm{r}) }{\partial \theta\partial  x} \nonumber \\
&&+ \sum_{m,n=-S}^S\sum_{\mu,\nu}^{\text{cgs}} 
\tilde P_{\mu(0)\nu(n)} \frac{\partial \tilde \chi^*_{\mu(m)}(\bm{r})}{\partial  x}  \frac{\partial \tilde \chi_{\nu(m+n)}(\bm{r}) }{\partial \theta} \nonumber\\
&&+  \sum_{m,n=-S}^S\sum_{\mu,\nu}^{\text{cgs}}
\tilde P_{\mu(0)\nu(n)} \frac{\partial \tilde \chi^*_{\mu(m)}(\bm{r})}{\partial \theta} \frac{\partial \tilde \chi_{\nu(m+n)}(\bm{r}) }{\partial  x}  \nonumber\\
&&+  \sum_{m,n=-S}^S \sum_{\mu,\nu}^{\text{cgs}}
\tilde P_{\mu(0)\nu(n)} \frac{ \partial^2 \tilde \chi^*_{\mu(m)}(\bm{r}) }{ \partial \theta\partial  x}  \tilde \chi_{\nu(m+n)}(\bm{r}) , \nonumber\\
\end{eqnarray}
\begin{eqnarray}
{ \frac{\partial^2 \rho(\bm{r})}{\partial \theta \partial y} }^{(b)} 
&=&\sum_{m,n=-S}^S \sum_{\mu,\nu}^{\text{cgs}} 
\tilde P_{\mu(0)\nu(n)} \tilde \chi^*_{\mu(m)}(\bm{r}) \frac{\partial^2 \tilde \chi_{\nu(m+n)}(\bm{r}) }{\partial \theta\partial  y} \nonumber\\
&&+ \sum_{m,n=-S}^S\sum_{\mu,\nu}^{\text{cgs}}
\tilde P_{\mu(0)\nu(n)} \frac{\partial \tilde \chi^*_{\mu(m)}(\bm{r})}{\partial  y}  \frac{\partial \tilde \chi_{\nu(m+n)}(\bm{r}) }{\partial \theta}\nonumber \\
&&+ \sum_{m,n=-S}^S\sum_{\mu,\nu}^{\text{cgs}} 
\tilde P_{\mu(0)\nu(n)} \frac{\partial \tilde \chi^*_{\mu(m)}(\bm{r})}{\partial \theta} \frac{\partial \tilde \chi_{\nu(m+n)}(\bm{r}) }{\partial  y}  \nonumber\\
&&+ \sum_{m,n=-S}^S\sum_{\mu,\nu}^{\text{cgs}} 
\tilde P_{\mu(0)\nu(n)} \frac{ \partial^2 \tilde \chi^*_{\mu(m)}(\bm{r}) }{ \partial \theta\partial  y}  \tilde \chi_{\nu(m+n)}(\bm{r}) ,\nonumber\\
\end{eqnarray}
and
\begin{eqnarray}
{ \frac{\partial^2 \rho(\bm{r})}{\partial \theta \partial z} }^{(b)} 
&=& \sum_{m,n=-S}^S\sum_{\mu,\nu}^{\text{cgs}} 
\tilde P_{\mu(0)\nu(n)} \tilde \chi^*_{\mu(m)}(\bm{r}) \frac{\partial^2 \tilde \chi_{\nu(m+n)}(\bm{r}) }{\partial \theta\partial  z} \nonumber\\
&&+ \sum_{m,n=-S}^S\sum_{\mu,\nu}^{\text{cgs}} 
\tilde P_{\mu(0)\nu(n)} \frac{\partial \tilde \chi^*_{\mu(m)}(\bm{r})}{\partial  z}  \frac{\partial \tilde \chi_{\nu(m+n)}(\bm{r}) }{\partial \theta} \nonumber\\
&&+ \sum_{m,n=-S}^S\sum_{\mu,\nu}^{\text{cgs}} 
\tilde P_{\mu(0)\nu(n)} \frac{\partial \tilde \chi^*_{\mu(m)}(\bm{r})}{\partial \theta} \frac{\partial \tilde \chi_{\nu(m+n)}(\bm{r}) }{\partial  z} \nonumber \\
&&+ \sum_{m,n=-S}^S\sum_{\mu,\nu}^{\text{cgs}} 
\tilde P_{\mu(0)\nu(n)} \frac{ \partial^2 \tilde \chi^*_{\mu(m)}(\bm{r}) }{ \partial \theta\partial  z}  \tilde \chi_{\nu(m+n)}(\bm{r}) .\nonumber\\
\end{eqnarray}
The first derivatives of the translating-rotating basis functions are given by Eqs.\ (\ref{chi_x})--(\ref{chi_z}) as well as by Eq.\ (\ref{chi_theta}), which 
are the equations actually programmed. 
The second derivatives are then evaluated as
\begin{eqnarray}
\frac{ \partial^2 \tilde \chi_{\mu(n)}(\bm{r}) }{ \partial \theta\partial  x} 
&=& \frac{ \partial^2 \tilde \chi_{\mu(n)}(\bm{r}) }{ \partial \theta\partial  \tilde x_{(n)}}, \\
\frac{ \partial^2 \tilde \chi_{\mu(n)}(\bm{r}) }{ \partial \theta\partial  y} 
&=& 
\frac{ \partial^2 \tilde \chi_{\mu(n)}(\bm{r}) }{ \partial \theta\partial  \tilde y_{(n)}} \cos n\theta 
%\nonumber \\&& 
- \frac{ \partial^2 \tilde \chi_{\mu(n)}(\bm{r}) }{ \partial \theta\partial  \tilde z_{(n)}} \sin n\theta 
\nonumber \\&& 
- \frac{ \partial \tilde \chi_{\mu(n)}(\bm{r}) }{\partial  \tilde y_{(n)}} n \sin n\theta 
%\nonumber \\&& 
- \frac{ \partial \tilde \chi_{\mu(n)}(\bm{r}) }{ \partial  \tilde z_{(n)}} n\cos n\theta, \nonumber\\ \\
\frac{ \partial^2 \tilde \chi_{\mu(n)}(\bm{r}) }{ \partial \theta\partial  z} 
&=& 
\frac{ \partial^2 \tilde \chi_{\mu(n)}(\bm{r}) }{ \partial \theta\partial  \tilde y_{(n)}} \sin n\theta 
%\nonumber \\&& 
+ \frac{ \partial^2 \tilde \chi_{\mu(n)}(\bm{r}) }{ \partial \theta\partial  \tilde z_{(n)}} \cos n\theta 
\nonumber \\&& 
+ \frac{ \partial \tilde \chi_{\mu(n)}(\bm{r}) }{\partial  \tilde y_{(n)}} n \cos n\theta 
%\nonumber \\&& 
- \frac{ \partial \tilde \chi_{\mu(n)}(\bm{r}) }{ \partial  \tilde z_{(n)}} n\sin n\theta, \nonumber\\
\end{eqnarray}
where Eqs.\ (\ref{tildex})--(\ref{tildez}) were used. The second derivatives in the right-hand sides are, in turn, given by
\begin{eqnarray}
\frac{ \partial^2 \tilde \chi_{\mu(n)}(\bm{r}) }{ \partial \theta\partial  \tilde x_{(n)}} 
&=& n\left( - y \sin n\theta +  z \cos n\theta  \right) \frac{\partial^2 \tilde \chi_{\mu(n)}(\bm{r}) }{\partial \tilde x_{(n)}\partial \tilde y_{(n)}}\nonumber  \\
&& + n\left( - y \cos n\theta -  z \sin n\theta  \right) \frac{\partial^2 \tilde \chi_{\mu(n)}(\bm{r}) }{\partial \tilde x_{(n)}\partial \tilde z_{(n)}} , \nonumber\\
\\
\frac{ \partial^2 \tilde \chi_{\mu(n)}(\bm{r}) }{ \partial \theta\partial  \tilde y_{(n)}} 
&=& n\left( - y \sin n\theta +  z \cos n\theta  \right) \frac{\partial^2 \tilde \chi_{\mu(n)}(\bm{r}) }{\partial \tilde y_{(n)}^2}  \nonumber\\
&& + n\left( - y \cos n\theta -  z \sin n\theta  \right) \frac{\partial^2 \tilde \chi_{\mu(n)}(\bm{r}) }{\partial \tilde y_{(n)}\partial \tilde z_{(n)}}, \nonumber \\ \\
\frac{ \partial^2 \tilde \chi_{\mu(m)}(\bm{r}) }{ \partial \theta\partial  \tilde z_{(n)}} 
&=& n\left( - y \sin n\theta +  z \cos n\theta  \right) \frac{\partial^2 \tilde \chi_{\mu(n)}(\bm{r}) }{\partial \tilde y_{(n)}\partial \tilde z_{(n)}}  \nonumber \\
&& + n\left( - y \cos n\theta -  z \sin n\theta  \right) \frac{\partial^2 \tilde \chi_{\mu(n)}(\bm{r}) }{\partial \tilde z_{(n)}^2}.
\end{eqnarray}
The second derivatives appearing in the right-hand sides of these equations are computed explicitly.

The basis function derivative of the multipole-expansion correction is
\begin{eqnarray}
\frac{\partial E_{\text{MPE}} }{\partial \theta}^{(b)} &=& \sum_{n=L+1}^{\infty} \frac{1}{(na)^3} \left\{ -4 \mu_x \frac{\partial \mu_x}{\partial \theta}^{(b)} 
+ 2 \mu_y \frac{\partial \mu_y}{\partial \theta}^{(b)} \cos n\theta 
\right. \nonumber\\&& \left.
+ 2  \mu_z \frac{\partial \mu_z}{\partial \theta}^{(b)} \cos n\theta 
- \mu_y^2 \,n \sin n\theta - \mu_z^2 \,n \sin n\theta \right\} . \nonumber\\
\end{eqnarray}
The derivative of the $y$ ``dipole moment'' is then given by
\begin{eqnarray}
\frac{\partial \mu_y}{\partial \theta}^{(b)} = \sum_{\mu,\nu}^{\text{cgs}} \sum_{n=-S}^S \left\{ \frac{\partial P_{\mu(0)\nu(n)} }{\partial \theta}^{(b)}  y_{\mu(0)\nu(n)}
+ P_{\mu(0)\nu(n)} \frac{\partial  y_{\mu(0)\nu(n)}}{\partial \theta}  \right\} , \nonumber\\
\end{eqnarray}
where the first derivative factor in the right-hand side is given by Eq.\ (\ref{Pder}) and the second derivative factor by 
\begin{eqnarray}
\frac{\partial   y_{\mu(0)\nu(n)}  }{\partial \theta} &=& 
\int d\bm{r}\,  \chi^*_{\mu(0)}(\bm{r}) y \frac{\partial  \chi_{\nu(n)}(\bm{r})}{\partial \theta} \\
&=& \int d\bm{r}\,  \chi^*_{\mu(0)}(\bm{r}) y
\nonumber\\&&\times\,
\left\{  \frac{\partial Y_{\nu(n)}}{\partial \theta}\frac{\partial  \chi_{\nu(n)}(\bm{r})}{\partial Y_{\nu(n)}} 
+  \frac{\partial Z_{\nu(n)}}{\partial \theta}\frac{\partial  \chi_{\nu(n)}(\bm{r})}{\partial Z_{\nu(n)}}  \right\} % \nonumber\\ 
\\
&=& (-n Y_{\nu(0)}\sin n\theta -n Z_{\nu(0)}\cos n\theta) 
\nonumber\\&&\times\, 
\int d\bm{r}\,  \chi^*_{\mu(0)}(\bm{r}) y 
 \left(-\frac{\partial  \chi_{\nu(n)}(\bm{r})}{\partial y} \right)  \nonumber \\
 && +  (n Y_{\nu(0)}\cos n\theta - n Z_{\nu(0)}\sin n\theta) 
\nonumber\\&&\times\, 
 \int d\bm{r}\,  \chi^*_{\mu(0)}(\bm{r}) y
 \left(-\frac{\partial  \chi_{\nu(n)}(\bm{r})}{\partial z} \right) .
\end{eqnarray}
It should be cautioned that these dipole derivatives would not agree with the numerical derivatives 
because the density matrix derivatives are not included in the former; Recall that Eq.\ (\ref{Pder}) only accounts
for the effects of basis function rotations on the density matrix.
Nevertheless, the total gradients are correct by virtue of the Pulay force [Eq.\ (\ref{Pulay})] that properly
takes into account the derivatives of the density matrix elements.

%\bibliography{library.bib}

\end{document}